\documentclass[10pt, a4paper, twocolumn, aps, pra, showpacs, longbibliography, nofootinbib,superscriptaddress]{revtex4-2}
\pdfoutput=1
\usepackage[utf8x]{inputenc}
\usepackage{ucs}
\usepackage{amsmath}
\usepackage{amsfonts}
\usepackage{amssymb}
\usepackage{mathtools}
\usepackage{makeidx}
\usepackage{cellspace,booktabs}
\usepackage{natbib}
\usepackage{lipsum}
\usepackage{bm}
\usepackage{bbm}
\usepackage{relsize}
\usepackage{hyperref}
\hypersetup{colorlinks = true, linkcolor=magenta,citecolor=blue, urlcolor=magenta, bookmarksnumbered =  true}

\usepackage{color}
\definecolor{light-gray}{gray}{0.55}

\usepackage{microtype}

\usepackage{graphicx}

\newcommand{\ssm}{\rm\scriptscriptstyle}

\begin{document}
\begin{abstract}	
Variational methods have proven to be excellent tools to
approximate ground states of complex many body Hamiltonians. Generic tools like neural networks are extremely
powerful, but their parameters are not necessarily physically motivated. Thus, an efficient parametrization of the wave-function can become challenging. In this letter we introduce a neural-network based variational ansatz that retains the flexibility of these generic methods while allowing for a tunability with respect to the relevant correlations governing the physics of the system. 
 We illustrate the success of this approach on topological, long-range correlated and frustrated models. Additionally, we introduce compatible variational optimization methods for exploration of low-lying excited states without symmetries that preserve the interpretability of the ansatz.

\end{abstract}

\date{\today}
\author{Agnes Valenti}
\affiliation{Institute for Theoretical Physics, ETH Zurich, CH-8093, Switzerland}
\author{Eliska Greplova}
\affiliation{Institute for Theoretical Physics, ETH Zurich, CH-8093, Switzerland}
\affiliation{Kavli Institute of Nanoscience, Delft University of Technology, 2600 GA Delft, The Netherlands}
\author{Netanel H. Lindner}
\affiliation{Physics Department, Technion, 3200003 Haifa, Israel}
\author{Sebastian D. Huber}
\affiliation{Institute for Theoretical Physics, ETH Zurich, CH-8093, Switzerland}

\title{Correlation-Enhanced Neural Networks as Interpretable Variational Quantum States}

\maketitle

Studying many-body systems beyond analytically solvable Hamiltonians is a formidable challenge due to the exponential growth of the Hilbert space size with the number of particles. While quantum Monte Carlo methods offer an unbiased solution to that challenge, they are not applicable to systems exhibiting the notorious sign problem \cite{ceperley1986quantum,becca2017quantum,troyer2005computational}. For models where numerically exact methods are not available, we can resort to variational methods: a clever parametrization of trial wave-functions allows to capitalize on physical intuition. This approach has led to scientific breakthroughs such as the Bardeen-Cooper-Schrieffer theory of superconductivity \cite{bardeen1957theory}. More sophisticated approaches were developed based on the same principle of using knowledge about the system's physics to obtain an accurate parametrization. Examples include Slater-Jastrow \cite{jastrow1955many, mitavs1994quantum} or Gutzwiller-projected wave functions \cite{gutzwiller1963effect,becca2017quantum,sheng2009spin,iqbal2011projected,gong2013phase,hu2015competing}. While these ansätze are able to represent also highly correlated states parametrized in an interpretable way, their simplicity comes with limited variational freedom.

Instead of starting from a system-specific wave function, other more universal approaches have been developed that rely on a generic parametrization of a sub-manifold of the Hilbert-space.
Typically, the size of the spanned sub-space is controlled via a tuning parameter determining the number of optimizable variables and thereby the wave-function accuracy. An example of this class are tensor network states, that span a submanifold of the Hilbert space determined by specific entanglement properties \cite{white1992density,vidal2003efficient,vidal2007entanglement,verstraete2006criticality}. More recently, neural-network based variational ansätze have been brought forward, providing a flexible wave-function ansatz not limited by entanglement or dimensionality \cite{carleo2017solving}. Proposed architectures include Restricted Boltzmann Machines (RBMs) \cite{carleo2017solving} or feed-forward (convolutional) neural networks \cite{saito2017solving,saito2018machine,liang2018solving, choo2018symmetries}. Neural network ansätze have been successfully applied to a range of different bosonic as well as fermionic systems \cite{gao2017efficient, nomura2017restricted, chen2018equivalence, glasser2018neural,liang2018solving,carleo2018constructing,pastori2019generalized, luo2019backflow,kochkov2018variational,ferrari2019neural,deng2017quantum, freitas2018neural, cai2018approximating,han2019solving, rocchetto2018learning}. 
While in principle highly expressive and efficiently trainable, previously proposed neural network architectures suffer from the fact that their parameters are often not physically motivated or interpretable, such that an efficient representation of the sought-after wave function is not ensured. As a consequence, an exponential number of parameters might be required to obtain high-accuracy wave-functions \cite{westerhout2020generalization}. Especially for applications such as the study of (topological) quantum phase transitions or identification of excited states, this exponential scaling represents a significant challenge.

Here we propose a novel variational ansatz that combines advantages of both physically motivated and generic variational methods while, at the same time, combating their limitations.
In particular, we design a neural network variational ansatz that is explicitly customizable to the form of the interactions in the considered system. We extend the energy functional describing an RBM by introducing coupling terms reflecting physical intuition. These coupling terms correspond to correlations governing the physics of the system under consideration.  
Inclusion of such correlators leads to a  significant increase in precision and flexibility of neural nets, while keeping the number of optimizable parameters minimal. In particular, this correlated RBM (cRBM) allows us to study topological phase transitions as well as long-range correlated models that are outside the reach of quantum Monte Carlo methods. We show that cRBMs can capture ground and low-energy excited states equally well: we formulate a variational approach to obtain excited states without symmetries that does not modify the structure of the ansatz and thus preserves the interpretablity of the variational wavefunction.

In this manuscript we demonstrate the power of the cRBM variational ansatz by providing a complete description of the phase diagram of a model showcasing topological transitions: Kitaev's toric model in the presence of magnetic fields \cite{kitaev2003fault, gottesman1997stabilizer, lidar2013quantum, gottesman1998theory, andersen2020repeated}.
We further demonstrate the interpretability of our ansatz by linking the improved accuracy of the correlation functions  to the customized physical extension of the energy functional on a long-range correlated model, the transverse field Ising model at criticality. Finally, we evaluate the performance of the ansatz on the antiferromagnetic Heisenberg model on a triangular lattice. In particular, we show that introducing coupling terms in the RBM ansatz can be used as a generic extension alternatively to increasing the hidden neuron density.

{\it Toric code model.} We explain the main properties of the cRBM ansatz on the perturbed toric code model with periodic boundary conditions \cite{kitaev2003fault} described by the Hamiltonian
 \begin{align}
H=-\sum \limits_{s} A_s -\sum \limits_{p} B_p +\vec{h} \cdot \sum \limits_{i} \vec{\sigma}_i,
\label{eq:Hamiltonian}
\end{align}
where $\vec{\sigma}_i$ denotes the Pauli matrices $\vec{\sigma}_i=(\sigma^x_i,\sigma^y_i,\sigma^z_i)$. The stabilizer operators $A_s=\prod_{i \in s} \sigma^x_i$ and $B_p=\prod_{i \in p} \sigma^z_i$ mutually commute. Vertices (plaquettes) of a square lattice are denoted by the subscripts $s$ ($p$) and $i$ runs over the edges, where spin-$1/2$ degrees of freedom are located. We apply a magnetic field $\vec{h}=(h_x,h_y,h_z)$ uniformly on each spin.

For $\vec{h}=0$, Hamiltonian~\eqref{eq:Hamiltonian} corresponds to the well-understood toric code, where in the ground state all operators $A_s$ and $B_p$ yield an eigenvalue $+1$ \cite{kitaev2003fault}. This phase possesses topological order characterized by a four-fold degenerate ground state on a torus. 


\begin{figure}[]
\centering
\includegraphics[scale=0.65]{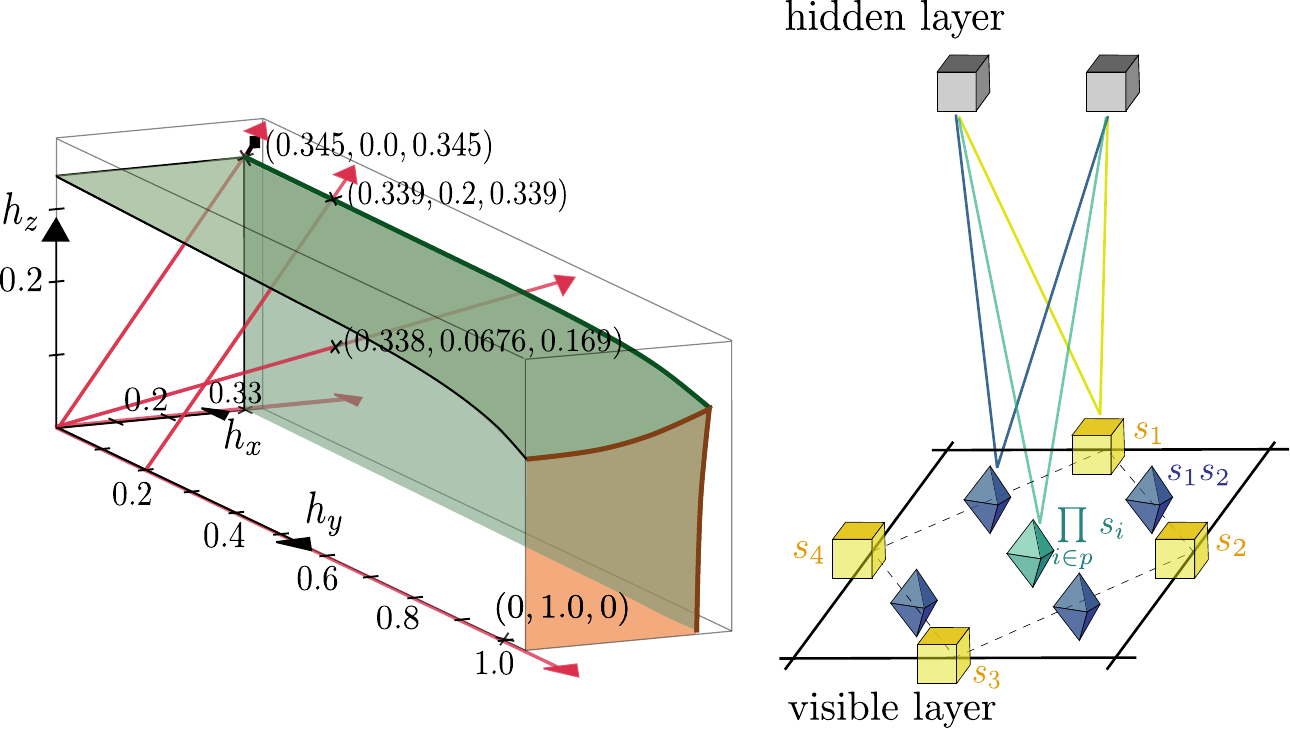}
\caption{Left: Conceptual phase diagram of the toric code. The red arrows show field directions in which we probed the diagram and the coordinates denote the obtained positions of the phase transition. The right panel shows the conceptual cRBM structure for a plaquette of the toric code model. Blue and green neurons and their connections to the hidden layer represent the added correlation terms.}
\label{fig:L1}
\end{figure}

The applied magnetic fields induce a phase transition out of the topologically protected phase. The nature and position of this transition depends on the direction of the applied field. While the   magnetic fields $h_x$ and $h_z$ are responsible for a second-order topological phase transition, a first-oder phase transition occurs when the transverse field $h_y$ dominates \cite{dusuel2011robustness,vidal2009self}.

The phase diagram (see Fig.~\ref{fig:L1}) has been previously explored by a variety of methods, but providing a unified approach capturing all features of this phase diagram has proven to represent a particular challenge. Numerically exact quantum Monte Carlo methods are only applicable to fields $\vec h \propto (h_x,0,h_z)$, i.e., parallel to the operators $A_s, B_p$ in (\ref{eq:Hamiltonian}) \cite{tupitsyn2010topological,wu2012phase}. One can circumvent this restriction by using either approximate methods such as advanced perturbation theory or by resorting to a specific class of variational wave functions in the form of infinite projected entangled pair states \cite{vidal2009low,vidal2009self,dusuel2011robustness}. In what follows, we show that the cRBM ansatz is able to capture the complete phase diagram.

{\it cRBM structure.} We introduce our proposed family of variational wave functions and highlight the differences to existing approaches. After exposing how to tailor the ansatz to a specific problem at hand we elaborate on how to determine the variational parameters. 

We consider a system of $N$ spin-$1/2$ degrees of freedom. In Ref. \onlinecite{carleo2017solving}, Carleo and collaborators have introduced RBMs as a variational ansatz for such a system
\begin{align}
\Psi(s_1, ...,s_N)&=\sum \limits_{\vec{\rho}}\exp[E_{\ssm RBM}(\Lambda)], \nonumber\\
E_{\ssm RBM}(\Lambda)&= \sum \limits_{k} a_k s_k\!+\!\sum \limits_{j} b_j  \rho_j\!+\!\sum \limits_{k,j} W_{k,j} s_k \rho_j, \label{eq:cRMB-terms}
\end{align}
where the $s_k \in \{-1,1\}$ represent the physical spins in a given basis and, in the language of RBMs, are termed the {\em visible layer}. The wavefunction ansatz $\Psi(s_1, ...,s_N)$ includes a sum over all possible values of the $M$ auxiliary spins $\vec{\rho}=(\rho_1, ...\rho_M)$, $\rho_i \in \{-1,1\}$, denoted as the {\em hidden layer}. The energy functional $E_{\ssm RBM}(\Lambda)$ can be understood as an interaction energy between classical spins. The parametrization $\Lambda=({\bf a},{\bf b},W)$ includes visible and hidden biases $a_k$ and $b_j$ as well as weights $W_{k,j}$ connecting the visible and hidden layer.

RBMs have been shown to represent exactly a wide class of topological states, including the unperturbed toric code ground state \cite{deng2017machine}. Here, we investigate to which extent modifications of RBMs are able to capture both extensions of the toric code as in (\ref{eq:Hamiltonian}) as well as systems where no exact RBM representation exists for any point of the phase diagram.

We modify the RBM ansatz by introducing {\em correlators to the energy functional} \eqref{eq:cRMB-terms}. We achieve this by adding visible neurons to the ansatz representing correlations between different spins. Concretely, the modification yields the energy functional
\begin{align}
E(\Lambda)= E_{\ssm RBM}(\Lambda)+\sum \limits_{i} a^{\rm corr}_{i} C_i
+\sum \limits_{i,j} W^{\rm corr}_{i,j} C_i \rho_j,
\label{eq:RBM_mod}
\end{align}
where $C_i=s_l...s_k$ are products between spins chosen to reflect dominant terms influencing system's behaviour. The sum $i$ runs over possible sets of spins entering the correlator $C_i$. These additional visible neurons are accompanied by their own biases and weights representing the interactions with the auxiliary spins of the hidden layer as shown in Fig.~\ref{fig:L1}.

The type of correlators $C_i=s_l...s_k$ that are added to the visible layer determine the efficiency of the use of the parameters in $\Lambda$. Moreover, it is in the design of these correlators, where the power of the cRMBs in allowing for the use of prior knowledge about the structure of the wavefunction can be capitalized on. In the spirit of conventional Jastrow-factor wave functions, the choice $C_i= s_l s_k$ representing two spin correlators can improve the wavefunction accuracy without an exponential growth in the number of hidden neuron density. Moreover, for specific problems we can include more complex terms. For example, in the case of the Hamiltonian (\ref{eq:Hamiltonian}) the relevance of plaquette operators can be mirrored on the level of the wavefunction by including toric code stabilizer operators $C_p=B_p=\prod_{i \in p} s_i$ in the energy functional. These extensions are depicted in Fig.~\ref{fig:L1}.  In order to further reduce the number of parameters and achieve higher precision, we symmetrize the ansatz \eqref{eq:RBM_mod} using translational symmetries as is usually done \cite{carleo2017solving}. The details of the ansatz and imposed symmetries are further specified in \cite{SI}.

\begin{figure}[]
\centering
\includegraphics[scale=1]{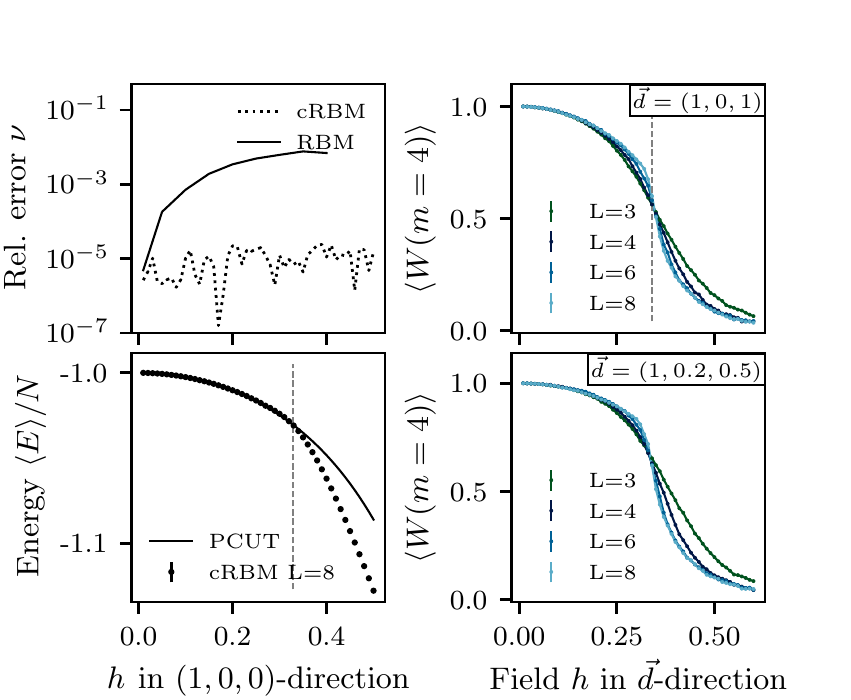}
\caption{Upper left:  relative error of the variational energy $\nu$ versus the field $h$ for $18$ spins. Lower left: the energy per spin for the cRBM ansatz for $128$ spins is shown together with the energy per spin obtained via PCUT. Upper right: Wilson loop expectation values $\langle W(m=4)\rangle$ on the self-dual line for different lattice sizes cross at the position of the phase transition, the literature value obtained via continuous-time Monte Carlo \cite{wu2012phase} is marked with a grey dashed line. Lower right: $\langle W(m=4)\rangle$ for fields in $(1,0.2,0.5)$-direction.}
\label{fig:L2}
\end{figure}

We fix the weights in $\Lambda$ by minimizing the cost function $\mathcal{C}$ corresponding to the variational energy
\begin{align}
\mathcal{C} \vcentcolon= \langle E \rangle =\frac{\langle \Psi | H| \Psi \rangle}{\langle \Psi | \Psi \rangle}
\end{align}
using stochastic reconfiguration, introduced by Sorella et al. \cite{sorella2007weak}. Expectation values are then calculated using variational Monte Carlo sampling \cite{mcmillan1965ground, sorella2007weak, park2020geometry, SI}. Computations using the standard RBM as comparison have been partly done with the help of the library {\it NetKet} \cite{carleo2019netket}.

{\it Results.} We now turn to the application of cRBMs to the investigation of the Hamiltonian (\ref{eq:Hamiltonian}). First, we assess the accuracy of our wavefunction by benchmarking it against exact results for small system sizes. In particular, we  emphasize the scaling of the accuracy with the number of variational parameters. To explore larger system sizes we compare our wavefunction with state-of-the art perturbation theory.

One can quantify the precision of the variational energy $E_{\ssm var}$ by comparing it with exact the value $E_{\ssm ED}$ obtained by exact diagonalization (ED) of Hamiltonian (\ref{eq:Hamiltonian}) on small lattices. We choose a lattice with $N=18$ spins, and compare the relative error $\nu=(E_{\ssm var}-E_{\ssm ED})/E_{\ssm ED}$ using a standard RBM ansatz with $\alpha=7$, where $\alpha$ corresponds to the number of hidden neurons. We compare it to an cRBM ansatz with the same amount of parameters (corresponding to $\alpha=2$). As shown in Fig. \ref{fig:L2}, $\nu$ of the cRBM ansatz is several orders of magnitude lower than the relative error obtained using the standard RBM ansatz. As shown in \cite{SI}, here the variational energy of the standard RBM ansatz cannot be significantly improved by increasing the hidden neuron density. Hence, the addition of the correlator terms in (\ref{eq:cRMB-terms}) allows to explore the relevant section of the Hilbert space more efficiently. 

We examine the performance of our ansatz for larger system sizes by comparing to state-of-the-art perturbation theory results using perturbative continuous unitary transformations
(PCUT) \cite{vidal2009low,dusuel2011robustness}. For a lattice with $N=128$ spins the variational and perturbative energies for the magnetic field $(h,0,0)$ are compared in Fig.~\ref{fig:L2}. The second-order phase transition for the chosen field direction is known to occur at $h_c=0.3284$ and the perturbation theory we compare to in Fig.~\ref{fig:L2} is know to be reliable for $h<h_c$. Figure~\ref{fig:L2} confirms that the variational energies match the perturbative energies well up to the second order phase transition, while yielding more accurate results outside of the topological phase. 

We can now use the cRBM ansatz to detect topological phase transitions. In particular, we probe the toric code in arbitrary field directions in order to recover the phase diagram, as depicted in Fig~\ref{fig:L1}. We are not limited to specific field directions as neither a sign problem occurs compared to Quantum Monte Carlo methods \cite{wu2012phase}, nor is the method restricted to a specific type of phase transition \cite{vidal2009low, dusuel2011robustness,castelnovo2008quantum}. We demonstrate our results on the self-dual line $(h,0,h)$ and a ray including a generic field directions $h(1,0.2,0.5)$. Recovering the characteristics of the self-dual line represents a particular challenge due to an occurring multicritical point and a first-order transition line outside of the topological phase \cite{tupitsyn2010topological, wu2012phase}. We detect the position of the topological phase transition in both cases using finite-size scaling analysis of the Wilson loop 
\begin{align}
\langle W(m) \rangle=\langle \prod^m \limits_{s} A_s \prod^m \limits_{p}B_p \rangle,
\end{align}
where $m$ determines the size of the loop.

Fig. \ref{fig:L2} shows the expectation value $\langle W(m=4) \rangle$ versus the magnitude of the applied field for system sizes up to $N=128$ spins. The curves for different system sizes cross approximately at the position of the phase transition which can be determined via a standard finite-size scaling analysis. We estimate the phase transition on the self-dual line $(h,0,h)$ to occur at $h\sim 0.345$, which is in accordance to literature values \cite{vidal2009low,wu2012phase}, see \cite{SI}. For the field direction $h(1,0.2,0.5)$, we obtain a transition at $h\sim 0.338$. This transition has not been determined in the literature previously. We confirm in \cite{SI} the found phase transitions by computing the {\it fidelity}, the overlap between two ground states with small difference in field strength that has been shown to scale to zero at the position of a second-order quantum phase transition \cite{zanardi2006ground,venuti2007quantum, hamma2008entanglement, valenti2019hamiltonian}. Long-range entanglement in the topological phase is found by probing the Renyi entanglement entropy \cite{hastings2010measuring}, see \cite{SI}.

\begin{figure}[]
\centering
\includegraphics[scale=1]{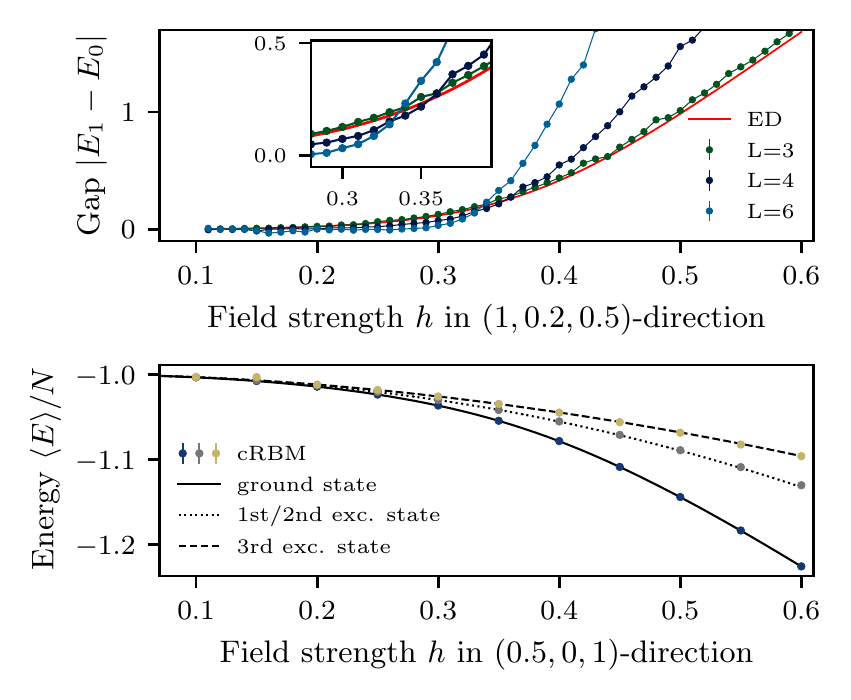}
\caption{Upper panel: The gap $|E_1-E_0|$ to the first excited state is shown for the lattice sizes $L=3,4,6$. For $L=3$, the gap obtained via ED is plotted as red line. The phase transition occurs at the crossing of the gaps for different lattice sizes (inset). 
Lower panel: The four lowest-lying ground states obtained via ED (lines) and with constrained cRBM (dots) for $L=3$, the three colors correspond to the different topological sectors.}
\label{fig:L3}
\end{figure}

{\it Excited states.} We have shown that our ansatz can represent the ground state of the studied topological model to high precision. However, relevant physics such as the splitting of topological degeneracies and excitations are encoded in the low-energy spectrum. Obtaining excited states with unknown quantum number poses a general challenge for variational wavefunctions \cite{choo2018symmetries}. We propose here a generic solution to this challenge that preserves the cRBM structure and flexibility of our ansatz.
In particular, we add the required orthogonality to the ground state we have previously found as a constraint to the cost function for the optimization of the wavefunction 
\begin{align}
\mathcal{C}=\frac{\langle \Psi | H| \Psi \rangle}{\langle \Psi | \Psi \rangle}+\kappa\frac{|\langle \Psi_0 | \Psi \rangle|^2}{\langle \Psi_0 | \Psi_0 \rangle \langle \Psi | \Psi \rangle},
\end{align}
where $|\Psi\rangle$ corresponds to the excited state to be optimized and $|\Psi_0\rangle$ is the ground state approximation determined in a previous step. The parameter $\kappa$ tunes the strength of the added constraint. The excited state is then obtained by minimizing the cost function using again stochastic reconfiguration. Extending the cost function while keeping the wave function ansatz intact allows to fully capitalize on the physically motivated ansatz as it preserves the wave function structure.

We compute the gap between the ground state and the first excited state for the field $h(1,0.2,0.5)$, as depicted in Fig. \ref{fig:L3}. The energy gap quantitatively matches the exact diagonalization result for small system sizes. Scaling of the energy gap unravels the topological degeneracy: the gap scales with system size as $e^{-L}$ (see \cite{SI}).

We introduce a further generalization of the cost function that allows for targeting specific states and does not rely on ground state ortogonalization. In particular, when considering a generic operator $M$, it is straightforward to obtain the state with lowest variational energy fulfilling the constraint $\langle M \rangle=A$ by minimizing the cost function
\begin{align}
\mathcal{C}=\frac{\langle \Psi | H| \Psi \rangle}{\langle \Psi | \Psi \rangle}+\kappa\bigg|\frac{\langle \Psi |M| \Psi\rangle}{\langle \Psi | \Psi \rangle} - A\bigg|^2
\label{eq:exc2}
\end{align}
for sufficiently large weight $\kappa$ and arbitrary constant $A$. This constrained minimization allows us to ``cherry-pick'' for an eigenstate with certain physical quantities. We can e.g. consider the four lowest-lying eigenstates of the perturbed toric model. Operators yielding different expectation values for the four states can straightforwardly identified as loops $\Gamma$ winding around the torus, as explained in \cite{SI}. Minimizing (\ref{eq:exc2}) with $M=\Gamma$ for a sufficiently wide range of $A$ and identifying eigenstates as the states with vanishing local energy variance or norm of variational energy derivatives \cite{becca2017quantum} allows to find all lowest-lying states of different topological sectors without the need to iteratively orthogonalize. Figure~\ref{fig:L3} depicts excited states found using this method and the cRBM ansatz for the field direction $(0.5,0,1)$. Higher excited states such as anyonic excitations \cite{kitaev2003fault} can in principle be obtained by suitable choice of operator $M$, e.g. stabilizers.

\begin{figure}[]
\centering
\includegraphics[scale=1]{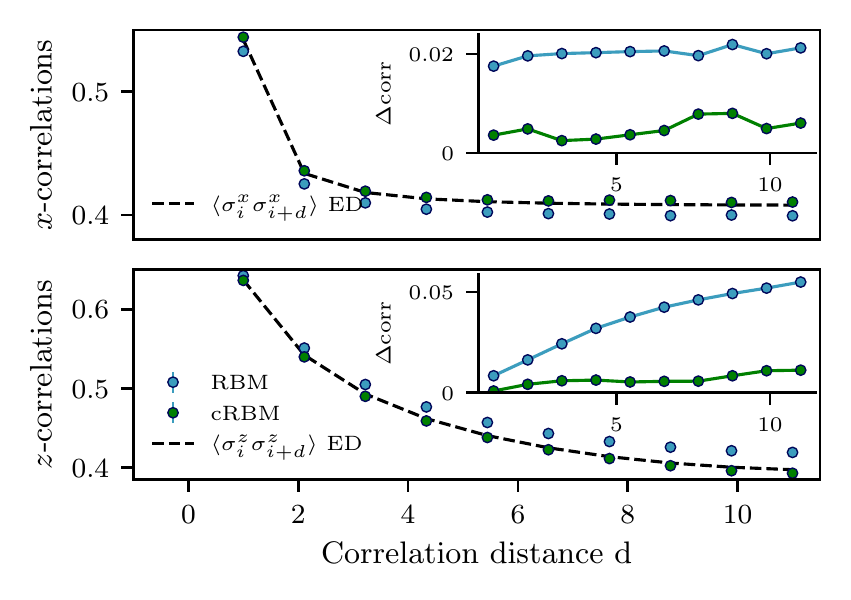}
\caption{1D TFI correlations: $z$-correlations and $x$-correlations obtained with an RBM (blue) and cRBM (green) ansatz are plotted as a function of the correlation distance for $N=22$ spins.}
\label{fig:L4}
\end{figure}

{\it Transverse-field Ising model.} We have shown on the example of the toric code, that introducing correlators tailored to the considered model to the energy functional increases the precision of the neural-network ansatz by several orders of magnitude. Additionally, let us demonstrate that the cRBM ansatz is also beneficial in a more general sense: we show that adding neurons representing generic correlations into the visible layer (as in Eq.~\eqref{eq:RBM_mod}) rather than the hidden layer leads to a more efficient approximation of the sought-after wave function. In particular, we compare the effect of adding simple correlators $C_{i,j}=s_i s_j$ (with nearest-neighbor spins $s_i$ and $s_j$) to the effect of increasing the hidden neuron density using the example of the transverse-field Ising model at the critical point \cite{sachdev2007quantum,chakrabarti2008quantum}.

The transverse-field Ising (TFI) model in 1D is described by the Hamiltonian
\begin{align}
H=-\sum \limits_{\langle i,j \rangle} \sigma^z_i \sigma^z_j - h \sum \limits_{i} \sigma^x_i,
\end{align}
where the sum runs over all nearest neighbors in the chain. The parameter $h$ tunes the strength of the transverse field. At the critical point $h=1$, the model exhibits long-range (algebraically decaying) spin-spin correlations \cite{sachdev2007quantum,chakrabarti2008quantum}.

In the standard RBM the spin-spin correlations are encoded in the sum over hidden neurons that forces the correlations to take a specific form (for details see \cite{SI}). Using the cRBM extension, we are able to encode the spin-spin correlations in the visible layer allowing for increased flexibility of their representation. We show that this flexible encoding captures long range correlations arising in the TFI model.

To make a simple comparison, we consider ansätze with a small hidden neuron density. In particular, we choose a standard symmetrized RBM ansatz\cite{carleo2017solving} with $\alpha=3$ hidden neurons and a cRBM with the same amount of parameters. Figure~\ref{fig:L4} shows the spin-spin correlation obtained for both variational wave functions. For both $x$-and $z$-correlations, the cRBM ansatz provides more accurate results. The most striking improvement appears for long-range $z$-correlations, where $z$ also corresponds to the choice of basis for the RBM (cRBM) ansatz. In \cite{SI} we illustrate the above mentioned improvement for the antiferromagnetic Heisenberg model on a triangular lattice 
\cite{leung1993spin,iqbal2016spin,capriotti1999Long,miyashita1984variational,huse1988simple,sheng2009spin,iqbal2011projected,gong2013phase,hu2015competing,schulz1996magnetic} and we show that the improved accuracy of the cRBM ansatz can be further extended to frustrated systems.

{\it Data availability.} In \cite{cm-cRBM}, we provide the code needed to calculate energies and topological quantities from a pre-trained cRBM wave-function.

{\it Perspectives.} We showed that a combination of generic neural-network based ans\"atze and available knowledge about the system can lead to significant improvements in precision of variational methods.
We introduced a correlator RBM ansatz with physically tunable flexibility and we demonstrated its power on topological, long-range correlated and frustrated models. In addition to formulating the cRBM ansatz we introduced compatible variational optimization techniques that allow for study of the low energy spectrum. An interesting additional pathway for further research is to examine how much further tunability is possible, e.g. by combining problem-specific neural-network architectures with information obtained via more conventional methods like perturbation theory, renormalization group, or mean-field methods \cite{ferrari2019neural}.


\section*{Acknowledgements}
We are thankful for enlightening discussions with Eyal Bairey and Giacomo Torlai.
We are grateful for financial support from the Swiss National Science Foundation, the NCCR QSIT. This work has received funding from the European Research Council under grant agreement no. 771503.

\vfill


\pagebreak
\clearpage
\widetext

\begin{center}
\textbf{\large Supplementary material: Correlation-Enhanced Neural Networks as Interpretable Variational Quantum States}
\end{center}

\setcounter{equation}{0}
\setcounter{figure}{0}
\setcounter{page}{1}
\makeatletter
\renewcommand{\theequation}{S\arabic{equation}}
\renewcommand{\thefigure}{S\arabic{figure}}
\section{Architecture details: Toric code}
\label{app:architecture}

\begin{figure}[b]
\centering
\includegraphics[scale=2.9]{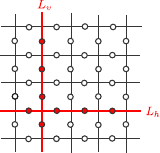}
\caption{Examples of vertical and horizontal non-contractible loops $L_v$ and $L_h$ are depicted here.}
\label{fig:loops}
\end{figure}

We explain and physically motivate the details of the architecture of the cRBM ansatz employed for the perturbed toric code. In particular, we begin by considering the toric code Hamiltonian (\ref{eq:Hamiltonian})
\begin{align}
H=-\sum \limits_{s} A_s -\sum \limits_{p} B_p,
\end{align}
where $A_s=\prod_{i \in s} \sigma^x_i$ and $B_p=\prod_{i \in p} \sigma^z_i$. When periodic boundary conditions are employed, the ground state (eigenstate of all stabilizer operators) is four-fold degenerate. In particular, the topological sectors differ by non-contractible loops: Let us consider the quantity $\Gamma=\prod_{i \in L} s_i$, where $L$ can be a horizontal or a vertical non-contractible loop as depicted in Fig.~\ref{fig:loops}. 
As $[C,H]=0$, the ground state manifold can be separated in four sectors with different eigenvalues $\{\lambda_v,\lambda_h \}$ of the vertical or horizontal loop. Since $\{\lambda_v,\lambda_h \}=\{\pm 1,\pm 1 \}$, there are four possible sectors corresponding to the four degenerate ground states.

In \cite{deng2017machine}, an exact RBM representation of the toric code ground state has been given. The RBM representation is constructed with a hidden layer, where each hidden neuron is associated with either a specific vertex or a specific plaquette. In particular, the neuron is then connected to the spins associated to the vertex $s$ or plaquette $p$  with the weights  $W_{s,i}=i\pi /2$ for $i \in s$ and $W_{p,i}=i\pi /4$ for $i \in p$, respectively. As there are in total as many stabilizers as physical spins, the hidden neuron density is equal to $1$. The toric code ground state is recovered when setting all biases equal to zero. The obtained ground state corresponds to an equal superposition of the four topological sectors (the sign of a specific sector in the superposition differs for different lattice sizes).

\begin{figure}[tb]
\centering
\includegraphics[scale=1]{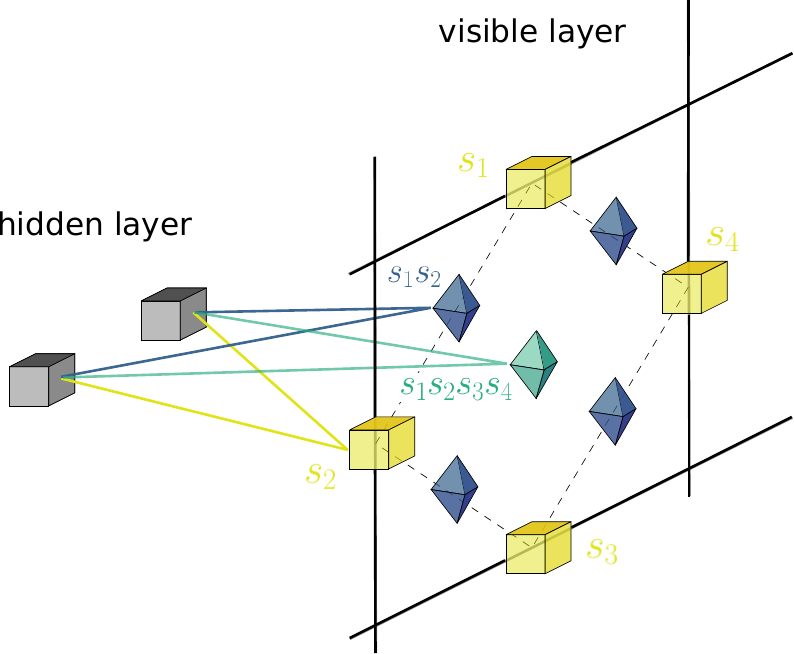}
\caption{The structure of the cRBM ansatz: In addition to the visible layer (yellow) an RBM would normally have, we add additional visible neurons (blue and green) capturing the relevant correlations in the system and connect them to the already existing hidden neurons (grey).}
\label{fig:ansatz}
\end{figure}

We symmetrize the ansatz by imposing translational symmetries. In the case of a toric code, the lattice can be separated into $A$-and $B$-lattice and therefore has a two-atomic basis. In particular, we recall that a standard RBM ansatz can be factorized to 
\begin{align}
\Psi(\{s_1, ...s_N\})=\exp(\sum \limits_{k} a_k s_k)\prod \limits_{j} \cosh (\sum \limits_k W_{k,j} s_k +b_j).
\end{align}
As a consequence, spin-spin correlations are encoded in the structure of the occuring cosine hyperbolicus. This property induces a restriction in the flexibility of the ansatz to represent correlations.
Let us denote the group of translational symmetries by $G$ and a group element by $g \in G$ defined by its action $g: k \to g(k)$ on a spin $k$. Then, the symmetrized ansatz is given by
\begin{align}
\Psi_{\rm symm}(\{s_1, ...s_N\}) =\exp(\sum \limits_{k} a_{{\rm bs}(k)} s_k)\prod \limits_{g \in G}\prod^{\alpha} \limits_{j=1} \cosh (\sum \limits_k W_{k,j} s_{g(k)} +b_j),
\end{align}
where ${\rm bs}(k)$ returns $0$ ($1$) if the spin $k$ is on $A$ ($B$)-sublattice. The parameter $\alpha$ denotes the number of hidden neurons. The toric code can be consequently written as symmetrized RBM with $\alpha=2$ and the weights
\begin{align}
W_{k,j}=\begin{cases} i\pi /2 &\text{if } j=0 \text{ and } k \in s_0, \\
i\pi /4 &\text{if } j=1 \text{ and } k \in p_0, \\
0 &\text{else.}
\end{cases}
\label{eq:toricweights}
\end{align}
Here, $s_0$ and $p_0$ correspond to an arbitrary vertex and plaquette. As the ansatz is symmetric, the choice of $s_0$ and $p_0$ does not matter.

The cRBM ansatz we choose includes generic correlators as well as correlators tailored to the toric code. In particular, we include
\begin{align}
C^{\rm bond}_{k,l}=s_k s_l, \ \ C^{\rm plaq}_{p}=\prod \limits_{k \in p} s_k, \ \ C^{\rm loop}_{L}=\prod \limits_{k \in L} s_k.
\end{align}
Here, $C^{\rm bond}_{k,l}$ corresponds to the product of nearest-neighbour spins. The correlator $C^{\rm plaq}_{p}$ consists of a product of all spins on the plaquette $p$, and non-contractible loops $L$ are included with $C^{\rm loop}_{L}$. The modifications are limited to the visible layer in order to preserve efficient sampling. The ansatz (only including bond- and plaquette correlators) is schematically depicted in Fig.~\ref{fig:ansatz}. Inserting the resulting energy functional into the symmetrized wave-function ansatz yields
\begin{align}
&\Psi_{{\rm cRBM},0}(\{s_1, ...s_N\})=\exp(\sum \limits_{k} a_{\rm bs}(k) s_k) \exp( \sum \limits_{\langle kl\rangle} a^{\rm bond}_{{\rm bsb}(k,l)} C^{\rm bond}_{k,l})\exp(\sum \limits_{p} a^{\rm plaq} C^{\rm plaq}_{p}) 
\exp(\sum \limits_{L} a^{\rm loop}_{{\rm bsL}(L)} C^{\rm loop}_{L}) \nonumber \\
&\times\prod \limits_{g \in G}\prod^{\alpha} \limits_{j=1} \cosh (b_j+\sum \limits_k W_{k,j} s_{g(k)} +\sum \limits_{\langle k,l\rangle} W^{\rm bond}_{kl,j}C^{\rm bond}_{g(k),g(l)} +\sum \limits_{p} W^{\rm plaq}_{p,j} C^{\rm plaq}_{g(p)} + \sum \limits_{L} W^{\rm loop}_{L,j} C^{\rm loop}_{g(L)}),
\label{eq:RBM_mod-app}
\end{align}
where we define the action of the symmetry operator $g$ on a set $M$ as the action of $g$ on all spins inside $M$, i.e. 
\begin{align}
C^{\rm plaq}_{g(p)}:=\prod \limits_{k \in p} s_{g(k)}, \ \ C^{\rm loop}_{g(L)}:=\prod \limits_{k \in L} s_{g(k)}.
\end{align}
In addition, there can be identified four distinct ``bond-sublattices'', that are left invariant by the applied translational symmetry. As a consequence, four biases associated to bond correlators remain when applying symmetries to the ansatz. The map ${\rm bsb}(k,l)$ returns the respective sublattice of the bond between site $k$ and $l$. As any plaquette can obtained by applying a suitable translational symmetry operation $g$ to another arbitrary plaquette, only one visible bias $a^{\rm plaq}$ associated to plaquette correlators exists. Concerning non-contractible loops, the map ${\rm bsL}(L)$ returns $0$ ($1$) if the loop is horizontal (vertical).

Additional flexibility is achieved by adding hidden neurons only connected to a certain type of visible neurons. Here, we improve the representability of superpositions of the topological sectors by extending the ansatz to
\begin{align*}
\Psi_{\rm cRBM}&=\Psi_{{\rm cRBM},0} \times \prod \limits_{g \in G} \cosh \big(b^{\rm vert}+\sum \limits_{L \in {\rm \mathcal{L}_{vert}}} W^{\rm vert. loop}_{L,j} C^{\rm loop}_{g(L)} \big) \times\cosh \big(b^{\rm horiz}+\sum \limits_{L \in {\rm \mathcal{L}_{horiz}}} W^{\rm horiz. loop}_{L,j} C^{\rm loop}_{g(L)} \big).
\end{align*}
The set of vertical (horizontal) non-contractible loops is denoted by ${\rm \mathcal{L}_{vert}}$ (${\rm \mathcal{L}_{horiz}}$). We have thus introduced two additional hidden neurons only connected to the visible neurons associated to non-contractible loops.

With chosen number of hidden neurons $\alpha$, the total number of parameters of our ansatz therefore corresponds to $11+2L+(1+2L+7L^2)\alpha$ on a lattice with length $L$. Throughout this manuscript, we choose $\alpha=2$.
For the conducted simulations, we initialize the ansatz in the toric code ground state. In addition, we add random small deviations on the parameters, as the pure toric code ground state (\ref{eq:toricweights}) represents a local minimum.

\section{Optimization and training}
\label{app:optimization}

\begin{figure}[tb]
\includegraphics[scale=1]{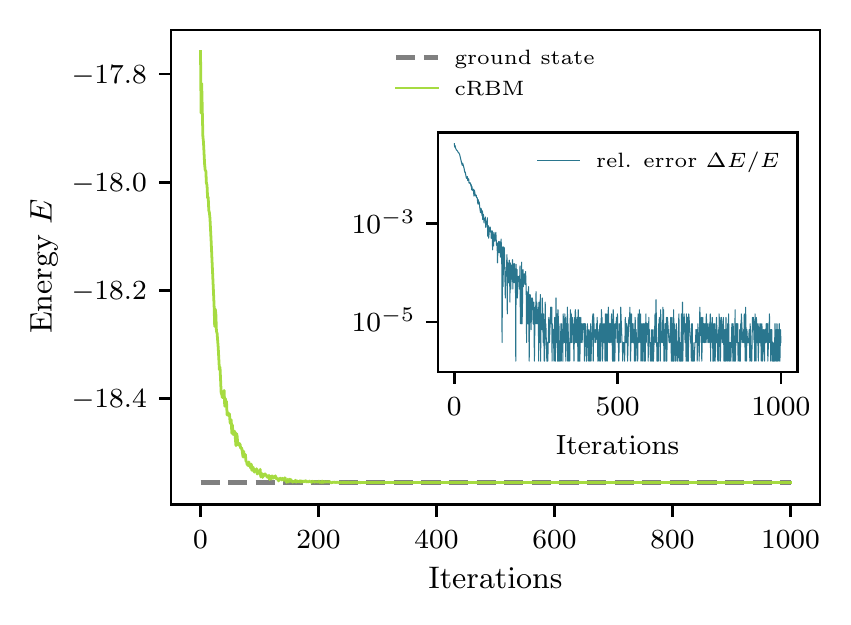}
\caption{Variational energies of a cRBM ansatz during the optimization procedure for the magnetic field $(0.3,0,0)$ on a lattice with $N=18$ spins. The ground state energy obtained via exact diagonalization is depicted as grey dashed line. The inset shows the relative error of the variational energy to the ground state energy with number of iterations.}
\label{fig:convergence}
\end{figure}

\subsection{Optimization}
We obtain the ground state wave function approximation by minimizing
\begin{align}
\langle E \rangle =\frac{\langle \Psi | H| \Psi \rangle}{\langle \Psi | \Psi \rangle}.
\end{align}
As optimization procedure, we implement the method introduced by Sorella et al. \cite{sorella2007weak}, which can be interpreted as effective second order approximation to the imaginary time evolution method: In every iteration step, the weights $\Lambda$ (representing all network parameters) are updated as
\begin{align}
\Lambda \to \Lambda - \eta S^{-1} \nabla_{\Lambda}\langle H \rangle.
\end{align}
Here, $\eta$ denotes the learning rate and second order effects are included by the covariance matrix \cite{park2020geometry}
\begin{align}
S_{k,k'}=\langle O^{*}_k O_{k'} \rangle -\langle O^{*}_k \rangle \langle O_{k'}\rangle.
\end{align}
The variational derivatives of the neural network ansatz $\Psi(\{s_1,...s_N\}):=\Psi(\mathcal{S})$ with respect to the $k$-th variational parameter $\Lambda_k$ are given by
\begin{align}
O_k(\mathcal{S})=\frac{1}{\Psi(S)}\partial_{\Lambda_k}\Psi(\mathcal{S}).
\end{align}
In addition, the force $\nabla_{\Lambda}\langle H \rangle$ can be reformulated as
\begin{align}
\nabla_{\Lambda_k}\langle H \rangle=\langle E_{loc} O^{*}_k\rangle -\langle E_{loc} \rangle \langle O^{*}_k \rangle,
\end{align}
where the local energy $E_{loc}$ is defined as
\begin{align}
E_{loc}(\mathcal{S})=\frac{\langle \mathcal{S}|H|\Psi \rangle}{\Psi(\mathcal{S})}.
\end{align}
In order to ensure that the inverse of the covariance matrix $S^{-1}$ is well-defined, we employ the explicit regularization $S=S+\epsilon id$. Throughout our work, we have typically chosen the learning rate $\eta$ in between $10^{-2}$ and $10^{-3}$, reducing the learning rate with number of iterations, and the regularization $\epsilon \sim 10^{-4}$. We note here, that it can be beneficial to decrease the regularization with number of iterations \cite{becca2017quantum, carleo2017solving}.
Fig.~\ref{fig:convergence} shows an exemplary convergence plot for the cRBM ansatz applied on the toric code with magnetic field $(0.3,0,0)$. We initialized the ansatz with the parameters specified in (\ref{eq:toricweights}) and additional small random noise.

\subsection{Efficient calculation of expectation values}
Expectation values of an operator $M$ with respect to the wave function ansatz $\Psi(\mathcal{S})$ can be calculated using variational Monte Carlo \cite{mcmillan1965ground, becca2017quantum}
\begin{align}
\frac{\langle \Psi | M | \Psi \rangle}{\langle \Psi | \Psi \rangle}=\frac{1}{\langle \Psi |\Psi\rangle}\sum \limits_{\mathcal{S},\mathcal{S}'} \Psi^{*}(\mathcal{S}) M(\mathcal{S},\mathcal{S}'),\Psi(\mathcal{S}')=\frac{1}{\langle \Psi |\Psi\rangle}\sum \limits_{\mathcal{S}} M_{loc}(S) \Psi^{*}(\mathcal{S})\Psi(\mathcal{S}),
\label{eq:vmc1}
\end{align}
where $M(\mathcal{S},\mathcal{S}')$ denotes the matrix element $\langle \mathcal{S} |M |\mathcal{S}'\rangle$ and
\begin{align}
M_{loc}(S)=\frac{\langle \mathcal{S}|M|\Psi\rangle}{\Psi(\mathcal{S})}.
\end{align}
Reformulating eq. (\ref{eq:vmc1}) as
\begin{align}
\frac{\langle \Psi | M | \Psi \rangle}{\langle \Psi | \Psi \rangle}=\frac{\sum \limits_{\mathcal{S}} M_{loc}(\mathcal{S})|\Psi(\mathcal{S})|^2}{\sum \limits_{\mathcal{S}}|\Psi(\mathcal{S})|^2},
\end{align}
we see that the expectation value can be efficiently calculated sampling the quantity $ M_{loc}(\mathcal{S})$ over a Markov Chain with update probability 
\begin{align}
p(\mathcal{S}_i \to \mathcal{S}_{i+1})={\rm min}\{\frac{|\Psi(\mathcal{S}_{i+1})|^2}{|\Psi(\mathcal{S}_{i})|^2}, 1\}.
\end{align}
For the toric model, we include vertex-flips and flipping of non-contractible loops in addition to single spin-flip updates in order to ensure ergodicity.

\subsection{Variational derivatives}
In order for the optimization described above to be feasible, it has to be possible to calculate the derivatives $O_k(\mathcal{S})$ efficiently. It has been shown in \cite{carleo2017solving}, that this statement holds for standard RBMs. We show here, that introducing correlators does not increase the complexity of the variational derivatives. Concretely, let us consider a generic cRBM ansatz of the form
\begin{align}
&\Psi(\mathcal{S})=\exp(\sum a_k s_k+\sum \limits_{i} a^{\rm corr}_{i} C_i)\prod \limits_{j} \cosh (\theta_j (\mathcal{S})),  \nonumber \\
&\theta_j(\mathcal{S})=\sum \limits_k W_{k,j} s_k+\sum \limits_{i} W^{\rm corr}_{i,j} C_i  +b_j.
\end{align}
The derivatives with respect to the network parameters are then gives by 
\begin{align}
&\frac{1}{\Psi(\mathcal{S})}\partial_{a_k} \Psi(\mathcal{S})=s_k, \\
&\frac{1}{\Psi(\mathcal{S})}\partial_{a^{\rm corr}_i} \Psi(\mathcal{S})=C_i, \\
&\frac{1}{\Psi(\mathcal{S})}\partial_{b_j} \Psi(\mathcal{S})=\tanh(\theta_j(\mathcal{S})), \\
&\frac{1}{\Psi(\mathcal{S})}\partial_{W_{k,j}} \Psi(\mathcal{S})=s_k\tanh(\theta_j(\mathcal{S})), \\
&\frac{1}{\Psi(\mathcal{S})}\partial_{W^{\rm corr}_{i,j} }\Psi(\mathcal{S})=C_i\tanh(\theta_j(\mathcal{S})).
\end{align}

\section{Accuracy of RBM and cRBM for the toric code}
\label{app:energies}
In this section, we compare in detail the accuracies of the RBM and cRBM wave functions for different field directions of the toric code.
With the purpose to understand the influence of the field directions to the accuracy of the wave-function ansatz, we examine the fields separately in $x$-, $y$- and $z$-direction on a small lattice with $N=18$ spins.
Simulations for the standard RBM were partly done with the help of the {\it netket} library \cite{carleo2019netket}.

\begin{figure}[tb]
\centering
\includegraphics[scale=1]{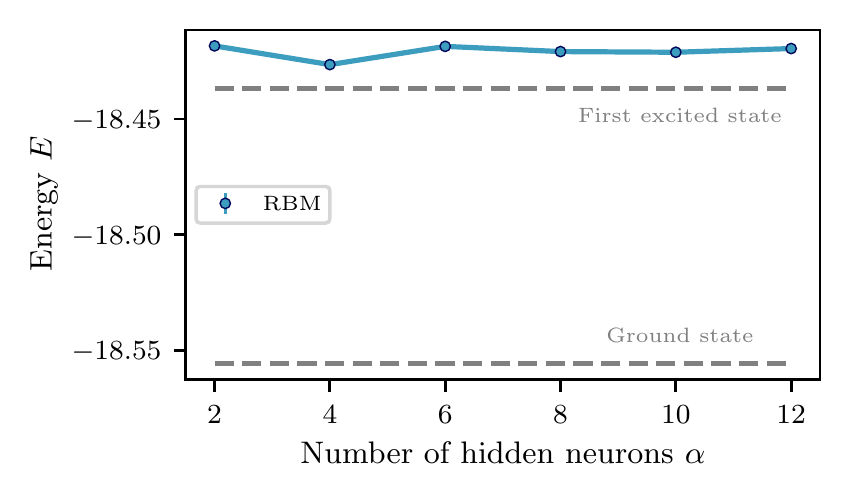}
\caption{RBM energies versus number of hidden neurons at $h_x=0.3$. The ground state and first excited state energy obtained using exact diagonalization are shown with grey dashed lines.}
\label{fig:alphahx}
\end{figure}

\paragraph{x-fields}
We have shown in the main text, that cRBMs achieve precisions several orders of magnitude higher than RBMs for the toric code with fields in $x$-direction (see Fig.~\ref{fig:L2}). In particular, we have compared an RBM and a cRBM with the same amount of parameters. In addition, increasing the hidden neuron density of the RBM ansatz does not lead to a significant increase in precision, as shown in Fig.~\ref{fig:alphahx} for $h_x=0.3$. Instead, as explained in the main text it represents a more relevant extension of the spanned sub-space to introduce correlators. 

\begin{figure}[b]
\includegraphics[scale=1]{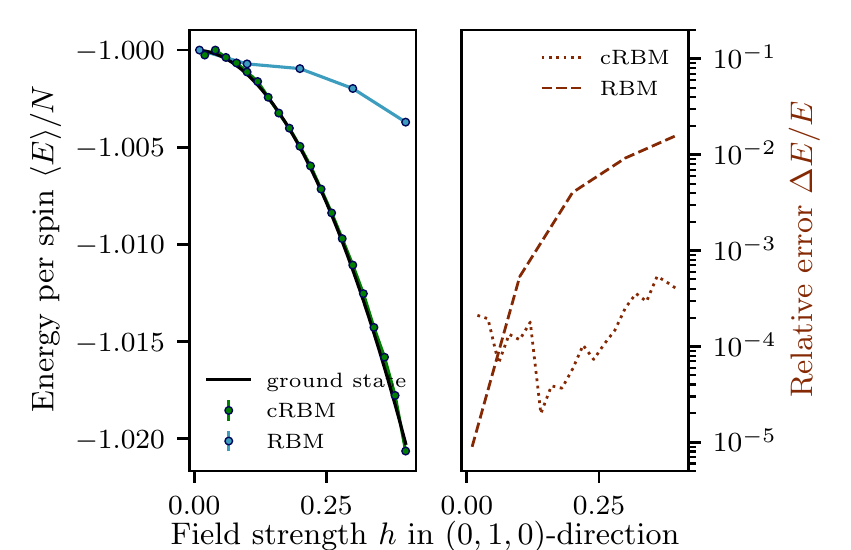}
\caption{RBM, cRBM energies for $(0,h_y,0)$ (left panel). The relative error with respect to the ground state energy obtained via exact diagonalization is plotted on the right hand side.}
\label{fig:energycomparehy}
\end{figure}

\begin{figure}[tb]
\includegraphics[scale=1]{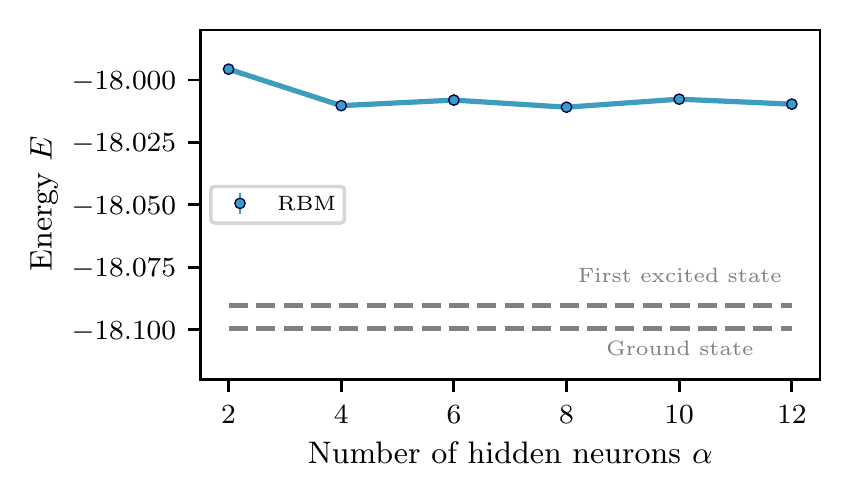}
\caption{RBM energies versus number of hidden neurons at $h_y=0.2$. The ground state and first excited state energy obtained using exact diagonalization are shown with grey dashed lines.}
\label{fig:alphahy2}
\end{figure}

\paragraph{y-fields}
We compare the energies of an RBM and a cRBM for fields in $(0,h_y,0)$-direction. As shown in Fig.~ \ref{fig:energycomparehy}, cRBMs again pose a significant improvement of the accuracy of the ansatz. It is in addition remarkable, that the RBM ansatz hardly learns any effect of the applied fields: For the ground state of the pure toric code $|\Psi_0\rangle$, the variational energy $\langle H \rangle$ is equal to $2L^2=18$ for any applied field. As we initialize the ansatz in the toric code ground state (plus small random deviations), $\langle H \rangle=18$ after optimization for nonzero $h_y$-fields corresponds to no net learning effect. As the ground state of the toric code with nonzero $h_y$ fields corresponds to a superposition of different topological sectors characterized by non-contractible loops, we deduce that including non-contractible loops as correlators mainly leads to the large improvement of the wave-function accuracy seen in Fig.~\ref{fig:energycomparehy} for the cRBM.
We tested again an increase of the hidden neuron density exemplary at $h_y=0.2$ for the RBM ansatz, and observed no significant improvement (see Fig.~\ref{fig:alphahy2}).

\paragraph{z-fields}
Let us consider the Hamiltonian
\begin{align}
H=-\sum \limits_{s} A_s - \sum \limits_{p} B_p + h_z \sum \limits_{i} \sigma^z_i.
\end{align}
For small $h_z$, we can approximate the Hamiltonian as \cite{castelnovo2008quantum}
\begin{align}
H_{M}&=-\sum \limits_{s} A_s - \sum \limits_{p} B_p+\sum \limits_{s} e^{h_z/2 \sum \limits_{i \in s} \sigma^z_i}\approx H+const.
\end{align}
The ground state of the model can be exactly represented as an RBM, starting from the exact representation of the toric ground state and assigning visible biases with values $a_i=-h_z$. As a consequence, an RBM ansatz yields a suitable approximation of the ground state of the original Hamiltonian $H$. Introducing additional correlators therefore yields only small improvements, if only $z$-fields are present. 
\section{Topological phases}
\label{app:topphase}

\begin{figure}[b]
\centering
\includegraphics[scale=1]{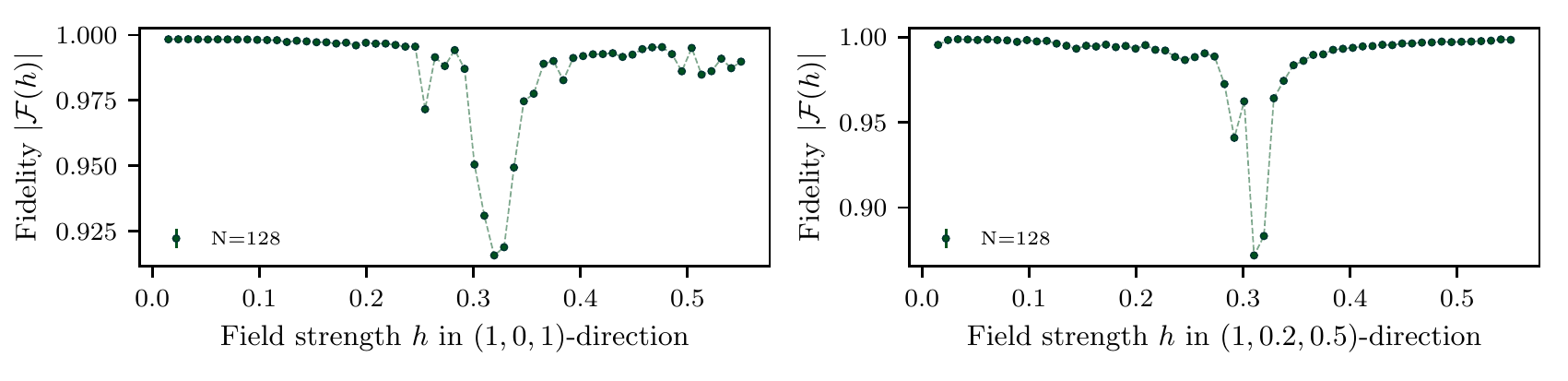}
\caption{Fidelities for the fields $(h,0,h)$ and $(h,0.2h,0.5h)$ for $N=128$ spins.}
\label{fig:fids}
\end{figure}

\begin{figure}[tb]
\centering
\includegraphics[scale=1]{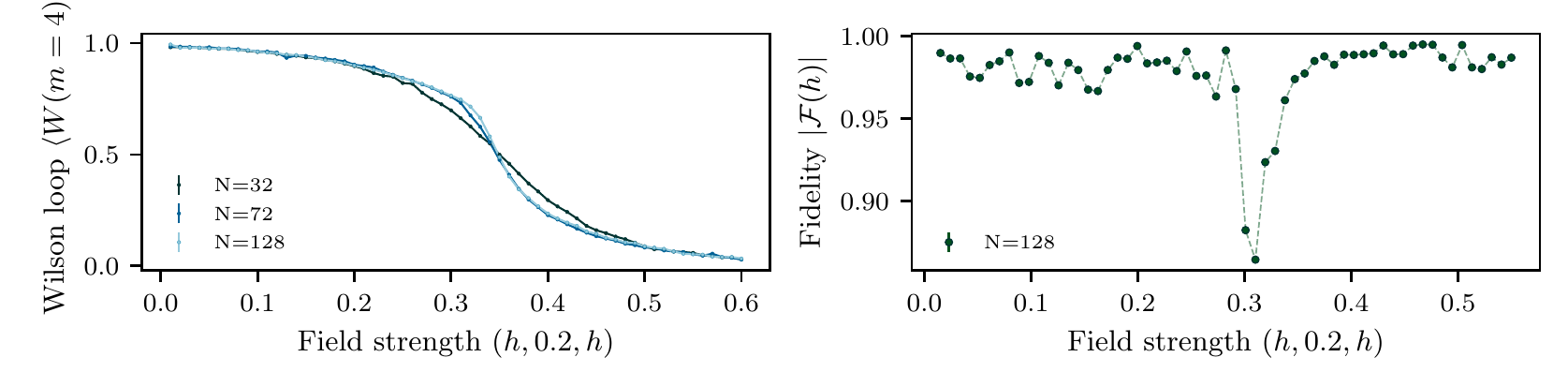}
\caption{Wilson loops and fidelity for the field direction $(h,0.2,h)$.}
\label{fig:fidloop}
\end{figure}

We probe the toric code phase diagram in different field directions and detect the topological phase transitions with a variety of methods. In the main text, we have examined the finite-size scaling behaviour of Wilson loops for selected field directions. We note here, as system sizes accessible to variational Monte Carlo methods are typically limited, finite size scaling analysis yields expectedly phase transition values slightly larger than the actual position (comparing e.g. the obtained transition around $h\sim 0.345$ on the self-dual line to literature values of $h\sim 0.340$ \cite{vidal2009low,wu2012phase}). In this section, we consider different methods and quantities in order to characterize the phase transitions and states in the respective phases further.

It has been shown, that the overlap between two ground states obtained at slightly varied field (``fidelity'') is a reliable tool to detect second-order symmetry-breaking quantum phase transitions \cite{zanardi2006ground,venuti2007quantum}. Numerical evidence suggests, that topological phase transitions can be detected as well by the fidelity \cite{hamma2008entanglement}. In particular, the fidelity is given by
\begin{align}
\mathcal{F}(h)=\frac{\langle \Psi_{h} | \Psi_{h+\Delta h} \rangle }{\sqrt{\langle \Psi_{h} | \Psi_{h} \rangle}\sqrt{\langle \Psi_{h+\Delta h} | \Psi_{h+\Delta h} \rangle} },
\label{eq:fid1}
\end{align}
where $\Psi_{h}$ denotes the ground state of a Hamiltonian $H(h)$.
At the position $h_C$ of the phase transition, the two states differ the most and the fidelity hence scales to zero.

We calculate the fidelity efficiently using a Monte Carlo Markov chain. In particular, we reformulate Eq.~(\ref{eq:fid1}) as
\begin{align}
&\mathcal{F}(h)=\frac{\sum \limits_{\mathcal{S}} \Psi^{*}_{h}(\mathcal{S}) \Psi_{h+\Delta h}(\mathcal{S})}{\sqrt{\sum \limits_{\mathcal{S}} \Psi^{*}_{h}(\mathcal{S}) \Psi_{h}(\mathcal{S})} \sqrt{\sum \limits_{\mathcal{S}} \Psi^{*}_{h+\Delta h}(\mathcal{S}) \Psi_{h+\Delta h}(\mathcal{S})} } =\frac{\sum \limits_{\mathcal{S}} \frac{\Psi_{h+\Delta h}(\mathcal{S})}{\Psi_{h}(\mathcal{S})}  |\Psi_{h}(\mathcal{S})|^2}{\sqrt{\sum \limits_{\mathcal{S}} 1 \cdot|\Psi_{h}(\mathcal{S})|^2} \sqrt{\sum \limits_{\mathcal{S}} \frac{|\Psi_{h+\Delta h}(\mathcal{S})|^2}{|\Psi_{h}(\mathcal{S})|^2}  |\Psi_{h}(\mathcal{S})|^2} }.
\label{eq:fid2}
\end{align}
A Markov chain with update probability $p(\mathcal{S}_{i} \to \mathcal{S}_{i+1})={\rm min}(1,|\Psi_{h}(\mathcal{S}_{i+1})|^2/|\Psi_{h}(\mathcal{S}_{i})|^2)$ can be used to calculate the quantities
\begin{align}
A_1=\frac{\sum \limits_{\mathcal{S}} \frac{\Psi_{h+\Delta h}(\mathcal{S})}{\Psi_{h}(\mathcal{S})}  |\Psi_{h}(\mathcal{S})|^2 }{\sum \limits_{\mathcal{S}} |\Psi_{h}(\mathcal{S})|^2}, \ \ \ A_2=\frac{\sum \limits_{\mathcal{S}} 1 \cdot  |\Psi_{h}(\mathcal{S})|^2 }{\sum \limits_{\mathcal{S}} |\Psi_{h}(\mathcal{S})|^2}=1, \ \ \ A_3=\frac{\sum \limits_{\mathcal{S}} \frac{|\Psi_{h+\Delta h}(\mathcal{S})|^2}{|\Psi_{h}(\mathcal{S})|^2}  |\Psi_{h}(\mathcal{S})|^2 }{\sum \limits_{\mathcal{S}} |\Psi_{h}(\mathcal{S})|^2}. \nonumber
\end{align}
The fidelity given in Eq.~\ref{eq:fid2} is then obtained as
\begin{align}
\mathcal{F}(h)=\frac{A_1}{\sqrt{A_2} \sqrt{A_3}}.
\end{align}
In order to account for a non-smooth gauge, we consider the norm of the fidelity $|\mathcal{F}(h)|$. Figure~\ref{fig:fids} shows the fidelity on the self-dual line for a lattice of $N=128$ spins as well as the fidelity for the field direction $h(1,0.2,0.5)$. The minimum in the fidelity indicates the respective second-order topological phase transition. We note here, that the fidelity scales to zero at the position of the phase transition with increasing system size. The critical field strenght is in accordance to the position of the phase transition obtained via finite-size scaling analysis of the quantity $W(m=4)$, see Fig.~\ref{fig:L2}.

We have probed the model for a variety of different field directions, exemplary discussing the self-dual line and the field direction $h(1,0.2,0.5)$. As another example, we show here the self-dual line for nonzero transverse field, i.e. $(h,0.2,h)$. The expectation values of the Wilson loop operator as well as the fidelity are depicted in Fig.~\ref{fig:fidloop}. The estimated critical field strength $h_C \sim 0.339$ is slightly smaller than the critical field strength for zero transverse field. This result is in agreement with the qualitative observation in Ref.~\cite{dusuel2011robustness}, that the critical field strength on the self-dual line decreases for increasing transverse field.

We further characterize the entanglement properties of the obtained states. In particular, we consider the generalized Renyi entanglement entropies
\begin{align}
S_n(\rho_A)=\frac{1}{1-n} \ln [{\rm Tr} (\rho^n_A)]
\end{align}
Here, $\rho_A$ is the reduced density matrix of the ground state on a subsystem $A$, obtained by tracing out a subsystem $B$. $A$ and $B$ correspond to a bipartition of the lattice. In the limit $n\to 1$, the van Neumann entanglement entropy is recovered
\begin{align}
S_1(\rho_A)=-{\rm Tr}(\rho_A \ln \rho_A).
\end{align}
The Renyi entropy for $n=2$ can be calculated efficiently via Variational Monte Carlo techniques. In particular, $S_2$ is given by 
\begin{align}
S_2=-\frac{1}{2}\ln \bigg[  \sum \limits_{\mathcal{S}_{A1},\mathcal{S}_{A2},\mathcal{S}_{B1},\mathcal{S}_{B2}} \frac{\Psi(\mathcal{S}_{A2}\cup \mathcal{S}_{B1})\Psi(\mathcal{S}_{A1}\cup \mathcal{S}_{B2})}{\Psi(\mathcal{S}_{A1}\cup\mathcal{S}_{B1})\Psi(\mathcal{S}_{A2}\cup\mathcal{S}_{B2})} |\Psi(\mathcal{S}_{A1}\cup \mathcal{S}_{B1}) \Psi(\mathcal{S}_{A2}\cup \mathcal{S}_{B2})|^2 \nonumber \\
\times \big( 1/ \sum \limits_{\mathcal{S}_{A1},\mathcal{S}_{A2},\mathcal{S}_{B1},\mathcal{S}_{B2}} |\Psi(\mathcal{S}_{A1}\cup \mathcal{S}_{B1}) \Psi(\mathcal{S}_{A2}\cup \mathcal{S}_{B2})|^2\big)\bigg].
\label{eq:S2}
\end{align}

Here, spin-configurations on sub-partition $A$ ($B$) are denoted with subscript $A$ ($B$). A spin-configuration on the whole lattice with the configuration on subpartition $A$ ($B$) corresponding to $\mathcal{S}_A$ ($\mathcal{S}_B$) coincides with $\mathcal{S}_A \cup \mathcal{S}_B$.
The second Renyi entropy can therefore be obtained by considering two copies of the lattice and the state $|\Psi\rangle \otimes |\Psi \rangle$. We construct a Markov chain with update probability 
\begin{align}
p(\mathcal{S}_i\to \mathcal{S}_{i+1})=\min (1, \frac{|\Psi(\mathcal{S}_{1,(i+1)}) \Psi(\mathcal{S}_{2,(i+1)})|^2}{|\Psi(\mathcal{S}_{1,(i)}) \Psi(\mathcal{S}_{2,(i)})|^2}),
\end{align}
where $\mathcal{S}_1$ ($\mathcal{S}_2$) corresponds to the spin-configuration on copy $1$ (copy $2$) of the lattice. In each step, either copy $1$ or copy $2$ is chosen and a spin-update proposed. Since the update probability factorizes, the efficiency of the computation is comparable to a standard Variational Monte Carlo calculation of any expectation value.
The obtained expression (\ref{eq:S2}) can be interpretated as the expectation value of a SWAP-operator, as detailed in Ref.~\cite{hastings2010measuring}.

We can derive expression (\ref{eq:S2}) by considering the reduced density matrix
\begin{align}
\rho_A=\sum \limits_{\mathcal{S}, \mathcal{S'}, \mathcal{S}_B} \langle \mathcal{S}_B|\mathcal{S} \rangle \langle \mathcal{S}'|\mathcal{S}_B\rangle \Psi^{*}(\mathcal{S})\Psi(\mathcal{S}')/\mathcal{N}^2,
\label{eq:rhoA}
\end{align}
where $\mathcal{N}$ denotes the normalization of the state $|\Psi\rangle$. The sum runs over all possible spin-configurations $\mathcal{S}$, $\mathcal{S}'$ on the whole lattice as well as all possible spin-configurations $\mathcal{S}_B$ on partition $B$. We reformulate expression~(\ref{eq:rhoA}) as
\begin{align}
\rho_A=\sum \limits_{\mathcal{S}_A, \mathcal{S}'_A, \mathcal{S}_B} \Psi^{*}(\mathcal{S}_A \cup \mathcal{S}_B) \Psi(\mathcal{S}'_A \cup \mathcal{S}_B) |\mathcal{S}_A \rangle \langle \mathcal{S}'_A|/\mathcal{N}^2, \nonumber
\end{align}
where $\mathcal{S}_A$ ($\mathcal{S}'_A$) denotes a spin-configuration on partition $A$.
Then,
\begin{align}
&{\rm Tr}(\rho^2_A) \nonumber \\
&=\frac{1}{\mathcal{N}^4} \sum \limits_{\mathcal{S}_A,\mathcal{S}_{A1},\mathcal{S}'_{A1},\mathcal{S}_{B1}}\langle \mathcal{S}_A |\mathcal{S}_{A1}\rangle \langle \mathcal{S}'_{A1} | \Psi^{*}(\mathcal{S}_{A1} \cup \mathcal{S}_{B1}) \Psi(\mathcal{S}'_{A1} \cup \mathcal{S}_{B1}) \sum \limits_{\mathcal{S}_{A2},\mathcal{S}'_{A2},\mathcal{S}_{B2}} |\mathcal{S}_{A2}\rangle \langle \mathcal{S}'_{A2} |\mathcal{S}_{A}\rangle \Psi^{*}(\mathcal{S}_{A2} \cup \mathcal{S}_{B2}) \Psi(\mathcal{S}'_{A2} \cup \mathcal{S}_{B2})  \nonumber \\
&=\frac{1}{\mathcal{N}^4}\sum \limits_{\mathcal{S}_{A1},\mathcal{A}_2,\mathcal{S}_{B1},\mathcal{S}_{B2}} \Psi^{*} (\mathcal{S}_{A1} \cup \mathcal{S}_{B1}) \Psi (\mathcal{S}_{A2} \cup \mathcal{S}_{B1}) \Psi^{*}(\mathcal{S}_{A2} \cup \mathcal{S}_{B2})\Psi (\mathcal{S}_{A1} \cup \mathcal{S}_{B2}). \nonumber 
\end{align}
Inserting the obtained expression together with 
\begin{align}
\mathcal{N}^4=\sum \limits_{\mathcal{S}_1} |\Psi(\mathcal{S}_1)|^2\sum \limits_{\mathcal{S}_2} |\Psi(\mathcal{S}_2)|^2 =\sum \limits_{\mathcal{S}_{A1},\mathcal{S}_{B1},\mathcal{S}_{A2},\mathcal{S}_{B2}} |\Psi(\mathcal{S}_{A1}\cup \mathcal{S}_{B1})  \Psi(\mathcal{S}_{A2}\cup\mathcal{S}_{B2})|^2
\end{align}
into the definition $S_2=-1/2 \ln[{\rm Tr} (\rho^2_A)]$ yields expression~(\ref{eq:S2}).

\begin{figure}[b]
\centering
\includegraphics[scale=1]{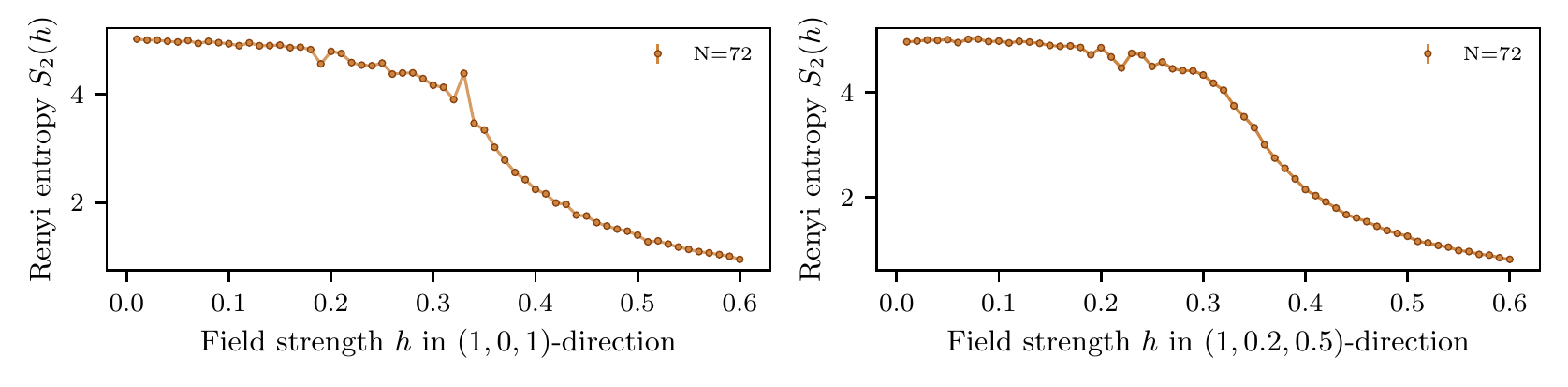}
\caption{Renyi entropies $S_2$ for  $(h,0,h)$ and $(h,0.2h,0.5h)$ for $N=72$ spins.}
\label{fig:S2}
\end{figure}

We have computed the Renyi entropy $S_2$ along the fields $(h,0,h)$ and $(h,0.2h,0.5h)$ for $N=72$ spins, as depicted in Fig.~\ref{fig:S2}. We observe, that the Renyi entanglement entropy is high in the topological phase, while approaching zero in the topologically trivial phase. This result is in accordance with long-range entanglement emerging as property of topological order. In addition, the calculation of the Renyi entropies explicitely shows, that the presented cRBM ansatz is able to capture the present long-range entanglement.

\section{Optimization with constraints}
\label{app:excstates}

\subsection{Excited states: Orthogonalizing}

\begin{figure}[tb]
\centering
\includegraphics[scale=1]{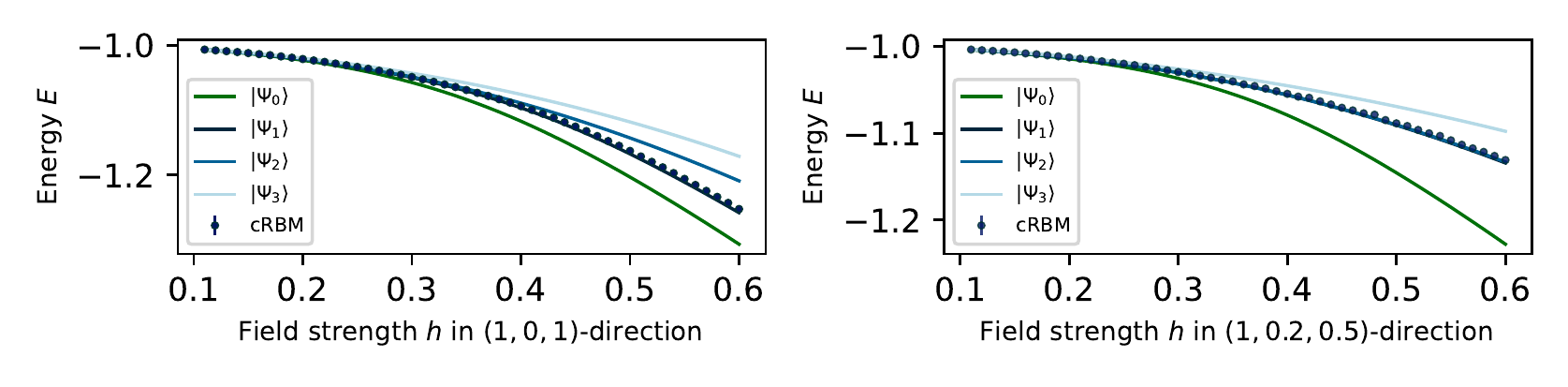}
\caption{The energies of the four lowest-lying states $|\Psi_0\rangle$, ..$|\Psi_3\rangle$ obtained via exact diagonalization on a lattice with $N=18$ spins are plotted for the field directions $(h,0,h)$ and $(h,0.2h,0.5h)$. The energies of the first excited state $|\Psi_1\rangle$ computed by optimizing a constrained cost function with a cRBM is in qualitative agreement with the exact diagonalization results.}
\label{fig:Exc}
\end{figure}

In this section, we detail the algorithm to obtain excited states presented in the main text. As explained in App.~\ref{app:optimization}, the ground state is obtained by minimizing the quantity (``cost function'')
\begin{align}
C(\Lambda):=\langle H \rangle.
\label{eq:C}
\end{align}
In order to find the first excited state, we extend the cost function by including the orthogonality constraint to the ground state. 
Given the ground state $\Psi_0$ obtained in a first variational optimization procedure, we define the modified cost function
\begin{align}
C_{E}(\Lambda)=\frac{\langle \Psi |H| \Psi \rangle}{\langle \Psi | \Psi \rangle}+\kappa\frac{|\langle \Psi_0|\Psi \rangle|^2}{\langle \Psi_0|\Psi_0\rangle \langle \Psi|\Psi \rangle}.
\end{align}
The parameter $\kappa$ tunes the weight of the constraint. More concretely, minimizing $C_{E}$ yields the state with lowest possible energy orthogonal to the ground state if $\kappa$ is sufficiently large. The minimization of the modified cost function is implemented straightforwardly. In particular, the modification appears in the computation of the derivative of the cost function in every natural gradient descent step.
\begin{align}
\Lambda \to \Lambda - \eta S^{-1} \nabla_{\Lambda}C_{E}.
\end{align}
Reformulating the cost function as
\begin{align}
C_{E}=\frac{\langle \Psi |H| \Psi \rangle}{\langle \Psi | \Psi \rangle}+\kappa \frac{\langle \Psi \big| P\big|\Psi \rangle} { \langle \Psi | \Psi \rangle} , \ \ P=\frac{|\Psi_0\rangle \langle \Psi_0|}{\langle \Psi_0 | \Psi_0 \rangle},
\end{align}
 the derivative with respect to the $k$-th network parameter is given by
\begin{align}
\partial_{\Lambda_k}C_{E}=\langle E_{loc} O^{*}_k \rangle -\langle E_{loc} \rangle \langle O^{*}_k \rangle +\kappa (\langle P_{loc}O^{*}_k \rangle -\langle P_{loc} \rangle \langle O^{*}_k \rangle).
\end{align}
Here, $P_{loc}$ is defined as
\begin{align}
P_{loc}(\mathcal{S})=\frac{\langle \mathcal{S}|P|\Psi \rangle}{\Psi(\mathcal{S})}.
\end{align}
The calculation of occuring expectation values containing $P_{loc}$ is in direct analogy to the calculation of the fidelity $\mathcal{F}$ discussed in Sec.~\ref{app:topphase}.

\begin{figure}[b]
\centering
\includegraphics[scale=1]{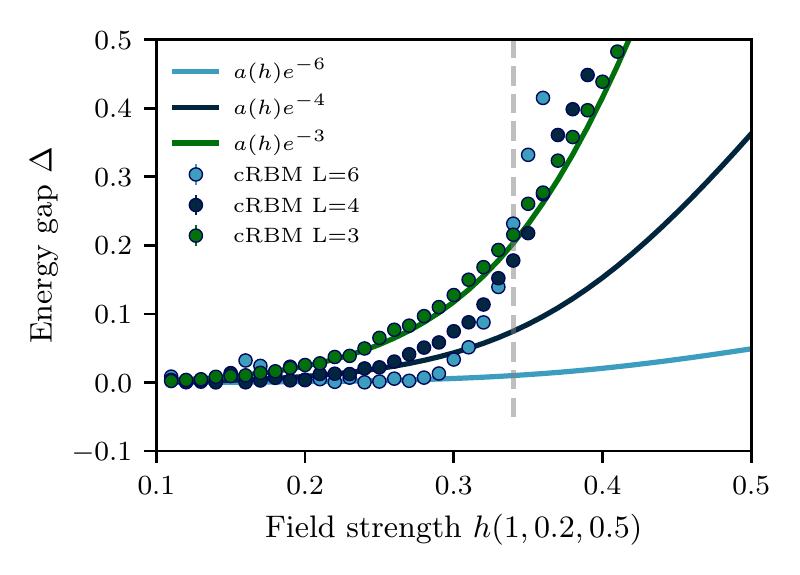}
\caption{The energy gap between the ground state and the first excited state obtained with a cRBM ansatz is plotted for different lattice sizes. The exponential scaling of the magnitude of the gap in the topological phase is plotted with solid lines. The position of the phase transition is indicated via a grey dashed line.}
\label{fig:topdeg}
\end{figure}

We demonstrate the accuracy of the received first excited state energies by comparison with exact diagonalization on a small lattice. More specifically, the comparison of the first excited state energy is depicted in Fig.~\ref{fig:Exc} for several field directions, showing quantitative agreement.

We have examined the scaling of the energy gap in the main text of this work. In particular, the magnitude of the energy gap $\Delta$ between the ground state and the first excited state has been shown to scale as  $\Delta \sim \exp (-L)$ in the topological phase of the toric model \cite{kitaev2003fault} yielding a degeneracy in the thermodynamic limit. We verify this scaling in Fig.~\ref{fig:topdeg}. 
In particular, we determine the proportionality constant $a(h)$ at field strength $\Delta(h) =a(h) \exp (-L)$ using exact diagonalization for a lattice of $N=18$ ($L=3$) spins. We plot $a(h) \exp (-L)$ together with the computed gap $\Delta(h)$ for each simulated lattice size and thus demonstrate the exponential decrease with system size of the energy gap in the topological phase.
The position of the phase transition coincides with the crossing of the energy gaps of different lattice size, as the decrease of the energy gap with increasing system size is a result of the ground state degeneracy in the thermodynamic limit, and thus only holds in the topological phase of the model.

\subsection{Excited states: Operator expectation values}
In this section, we examine the variational optimization procedure in the presence of constraints on the wave function. In particular, we consider a (hermitian) operator $M$ and aim to find the state with lowest variational energy fulfilling the constraint $\langle M \rangle=A$, where A is a (real) value. It has been shown in the main text, that such an optimization can be used to ``cherry-pick'' for an excited state with approximately known physical quantities. The procedure to obtain specific excited states consists of two steps: (1) determining the appropriate operator. Here, it is crucial that the operator is chosen such that no state with lower energy than the state of interest has a larger expectation value than the sought-after state (or, if states with larger expectation values and lower energies exist, no state with lower expectation value and lower energy should exist). This condition ensures, that indeed an eigenstate and not a superposition of eigenstates is obtained. (2) Optimizing for a large enough {\it range} of different values of $A$. The eigenstate(s) can be identified by considering the variance of the local energy or the norm of the local energy derivatives $||\nabla \langle H \rangle||$. Both quantities vanish for an eigenstate \cite{becca2017quantum}. 

\begin{figure}[bt]
\centering{
\includegraphics[scale=1]{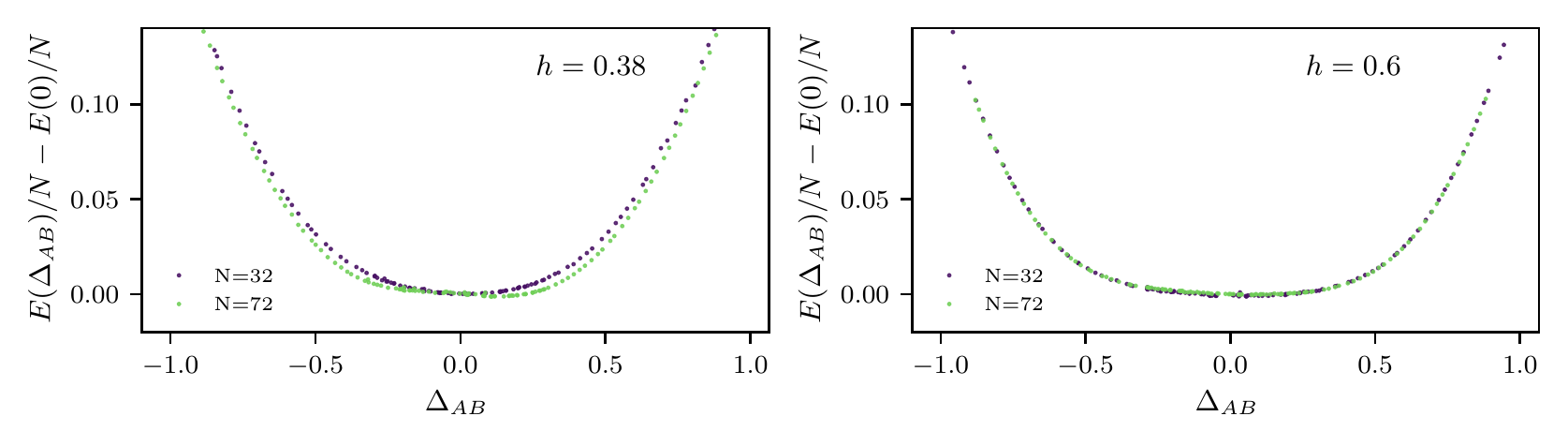}}
\caption{The difference $E(\Delta_{AB})/N-E(0)/N$ of the variational energy constrained to $\Delta_{AB}=\langle A_s-B_p \rangle$ for the field strenghts $h=0.38$ (upper panel) and $h=0.6$ (lower panel) for two lattice sizes. On the first order transition line, the difference decreases with larger lattice size, in particular in the range ($-0.5 \leq \Delta_{AB} \leq 0.5$) indicating the values of the parameter $\Delta_{AB}$ in the two adjacent phases.}
\label{fig:enconstrained}
\end{figure}

A constrained optimization might also be useful for the consideration of first-order phase transitions, choosing $M$ to be the respective order parameter and therefore being able to identify an occuring level crossing.

We here detail the optimization procedure. The ground state in general is obtained by minimizing the cost function $C(\Lambda)$ (\ref{eq:C}).
In order to impose the given constraint, we extend the cost function. In particular, we consider the cost function of the form
\begin{align}
C_{M}(\Lambda):=\langle H \rangle+\kappa |\langle M \rangle-A|^2
\end{align}
The added constraint $|\langle M \rangle-A|^2$ is minimal, when the expectation value of the hermitian operator $M$ is equal to the real value $A$.
The parameter $\kappa$ tunes the weight of the constraint. More concretely, minimizing $C_{\Lambda}$ yields the state with lowest possible energy fulfilling the constraint $\langle M \rangle = A$ if $\kappa$ is sufficiently large. The optimization algorithm is straightforwardly adapted. More concretely, the weights are updated in each step by
\begin{align}
\Lambda \to \Lambda - \eta S^{-1} \nabla_{\Lambda}C_{M}.
\end{align}
The derivative of the cost function with respect to the $k$-th parameter is given as
\begin{align}
\partial_{\Lambda_k}C_{M}=\partial_{\Lambda_k}\bigg( \frac{\langle \Psi| H | \Psi \rangle}{\langle \Psi| \Psi \rangle}+\kappa \big|\frac{\langle \Psi | M| \Psi \rangle}{\langle \Psi | \Psi \rangle}-A\big|^2\bigg) \nonumber \\
=\langle E_{loc} O^{*}_k \rangle -\langle E_{loc} \rangle \langle O^{*}_k \rangle +2\kappa (\langle M_{loc} \rangle -A)(\langle M_{loc} O^{*}_k \rangle -\langle M_{loc} \rangle \langle O^{*}_k \rangle).
\label{eq:derivC}
\end{align}
Here, $M_{loc}$ is defined as
\begin{align}
M_{loc}(\mathcal{S})=\frac{\langle \mathcal{S}|M|\Psi \rangle}{\Psi(\mathcal{S})}.
\end{align}
Updating the gradient descent step accordingly to Eq.~(\ref{eq:derivC}) thus results in a minimization of the generalized cost function $C_M$.

In the main text, we have shown how the procedure can be used to obtain excited states.
We here additionally use the constrained optimization procedure to characterize a first order quantum phase transition. In particular, we examine the first-order transition line on the self-dual line $h(1,0,1)$ of the toric code. Previous studies have shown, that a first order transition from the vortex condensed to the charge condensed phase occurs on the self dual line in the regime $0.34<h< 0.418$ \cite{wu2012phase, tupitsyn2010topological}. We examine the field strengths $h=0.39$ and $h=0.6$ together with the expectation value $\Delta_{AB}:=\langle A_s -B_p \rangle$. In the charge-condensed phase, $ \Delta_{AB}<0$ whereas $ \Delta_{AB}>0$ in the vortex-condensed phase. The quantity thus serves as parameter defining the present phase. A first-order transition is characterized by a phase coexistence. In particular, we here expect in the thermodynamic limit a ground state degeneracy of two states with different value of $\Delta_{AB}$. However, for finite lattice size a different (symmetric) state with $\Delta_{AB}=0$ might have a lower energy than the degeneracy of the two states with $\Delta_{AB}\neq 0$. When optimizing at $h=0.39$ with the constraint $\Delta_{AB}\neq 0$, we expect the difference of the obtained variational energy to the ground state energy to decrease for larger lattice sizes such that degeneracy arises in the thermodynamic limit. As an exemplary illustration, we plot the difference $E(A)/N-E(0)/N$ for two lattice sizes ($N=32$ and $N=72$ spins) for $-1\leq A \leq 1$. Here, $E(A)$ is the variational energy optimized under the constraint $\Delta_{AB}=A$. Figure~\ref{fig:enconstrained} shows $E(A)/N-E(0)/N$ for the cases $h=0.38$ and $h=0.6$. The difference indeed decreases (for $\Delta_{AB}$ in a certain range) on the first-order transition line ($h=0.38$), whereas no change in energy difference is observed outside of the first-order transition line.

\section{Heisenberg model on a triangular lattice}
\label{app:Heisenberg}

\begin{figure}[tb]
\centering
\includegraphics[scale=0.9]{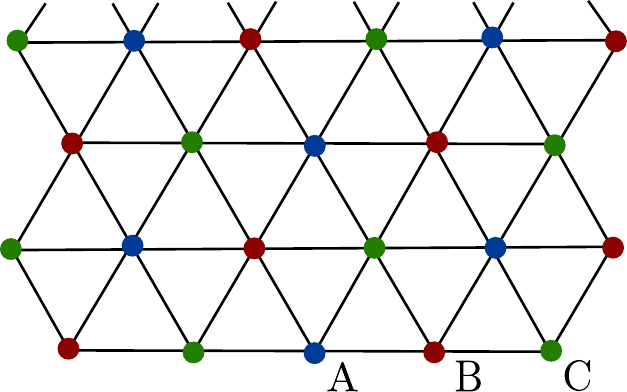}
\caption{Triangular lattice, A-,B- and C-sublattices are indicated in blue, red and green.}
\label{fig:lattice_triangular}
\end{figure}

\begin{figure}[b]
\centering
\includegraphics[scale=1]{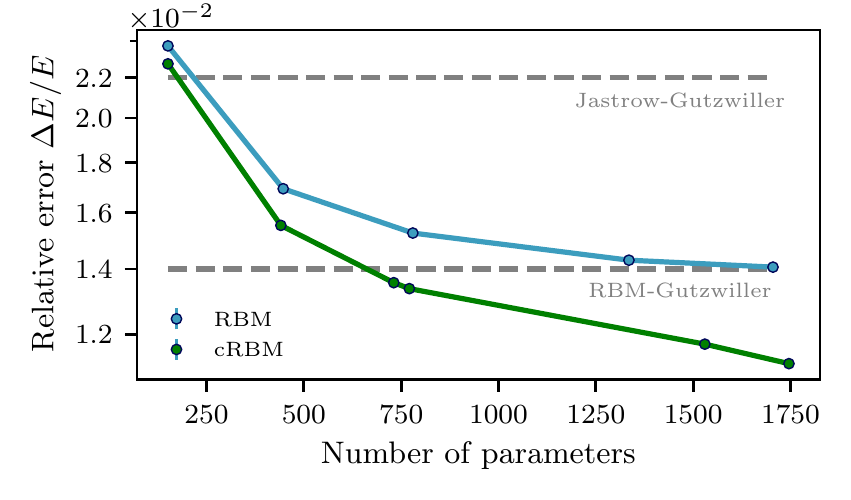}
\caption{Relative error of the variational energies obtained with an RBM and cRBM anatz for the triangular Heisenberg model on a model with $36$ spins.}
\label{fig:L5}
\end{figure}

We employ the cRBM ansatz for an example of a frustrated model, the antiferromagnetic Heisenberg model on a triangular lattice.

The Hamiltonian is given by
\begin{align}
H=J \sum \limits_{\langle i, j \rangle} \vec{\sigma}_i \cdot \vec{\sigma}_j,
\end{align}
where $J\geq 0$ is the antiferromagnetic interaction strength and $\vec{\sigma}=(\sigma^x,\sigma^y,\sigma^z)$. The sum is conducted over nearest-neighbors in the triangular lattice. Periodic boundary conditions as detailed in \cite{leung1993spin} are employed. Obtaining the ground state wave function has been proven to be a formidable challenge due to the Quantum Monte Carlo sign problem induced by the frustration in the system. In addition, variational ansätze have to overcome the difficulty to find the nontrivial sign structure of the real ground state wave function \cite{iqbal2016spin}.
The most accurate approximations of the ground state energy include the Green’s function Monte Carlo (GFMC) result of \cite{capriotti1999Long} and Gutzwiller-projected wave functions with Jastrow or RBM factor \cite{ferrari2019neural}.

We compare the energies obtained using a standard RBM on a 6x6-cluster with the energies obtained using the modified cRBM ansatz. In order to achieve better accuracies, the RBM is initialized in the $120^{\circ}$ Neel state. This classical state can be written as RBM by using the expression introduced in \cite{miyashita1984variational} and \cite{huse1988simple}:
\begin{align}
| \Psi \rangle_{\rm Neel}=\sum \limits_{s} \exp (\tilde{H}_0) |s\rangle, \ \ \ \ \tilde{H}_0=\frac{1}{3}\pi i (\sum \limits_{i \in B} s_i-\sum \limits_{i \in C} s_i)
\end{align}
with $|s\rangle$ a basis state in spin-$z$-basis and $s_i \in \{-1,1\}$ the value of the spin in $z$-direction at lattice site $i$. The sublattices are denoted with $A$, $B$ and $C$ as illustrated in Fig.~\ref{fig:lattice_triangular}.
As a consequence, $| \Psi \rangle_{\rm Neel}$ corresponds to an RBM with the parameters $b_j=0$, $W_{kj}=0$ and visible biases 
\begin{align*}
a_k=\begin{cases} 1/3 \pi i &\text{if } k \in $B$, \\
-1/3 \pi i &\text{if } k \in $C$, \\
0 &\text{else.}
\end{cases} 
\end{align*}
We employ symmetries on the RBM ansatz compatible with the symmetries of the Neel state and the periodic boundary conditions detailed in \cite{schulz1996magnetic}: translational symmetry which keeps the sublattice invariant and $C_{3v}$ symmetry. 
Due to the $SU(2)$ symmetry of the Heisenberg Hamiltonian, a more accurate ground state wave function is obtained by restricting to the subspace with $S_z=0$. This constraint is enforced by restricting the Monte Carlo sampling to configurations with total $z$-magnetization $\sum s_z=0$, only allowing (nearest-neighbor) spin-exchange updates.
The relative errors of the variational energies of the RBM ansatz are plotted in Fig.~\ref{fig:L5}. As parameters (hidden neurons) are increased, similar energies as the Gutzwiller-RBM ansatz are obtained. We perform the simulations on a $6\times 6$- lattice and compare with exact diagonalization \cite{schulz1996magnetic,capriotti1999Long}.

We compare the variational energies obtained with the RBM ansatz with a cRBM ansatz. The cRBM ansatz includes physical as well as generic extensions. In particular, we make use of the a correction introduced to the Neel state by Huse and Elser \cite{huse1988simple}
\begin{align}
| \Psi \rangle_{\rm Neel}=\sum \limits_{s} \exp (\tilde{H}_0+\tilde{H}_1 + \tilde{H}_2) |s\rangle, \\
\tilde{H}_1=-\sum \limits_{i,j} K_{i,j} s_i s_j, \ \ \ \tilde{H}_2=i L \sum \limits_{i,j,k} \gamma_{ijk} s_i s_j s_k,
\end{align}
where $\tilde{H}_1$ and $\tilde{H}_2$ correspond to two-body and three-body corrections, respectively. In particular, $\exp (\tilde{H}_1)$ takes the form of the Jastrow factor.  The sum in the expression for $\tilde{H}_1$ runs over nearest neighbors, while the sum in $\tilde{H}_2$ runs over distinct triplets: the sites $i$ and $k$ are second neighbours to one another while being both nearest neighbours of $j$. The parameters $K_{i,j}$ and $L$ determine the accuracy of the Huse-Elser wave function. The value of the sign factor $\gamma_{ijk}=\gamma_{kji}=\pm 1$ changes sign under rotations by $\pi/3$ or $\pi$. The corrections to the Neel state induce a significant improvement in the wave function accuracy and consitute an early and simple approach to a variational wave function for the triangular Heisenberg antiferromagnet. 

The Huse-Elser corrections can be naturally implemented as cRBM ansatz. In particular, we introduce $2$-body and $3$-body correlators as in $\tilde{H}_1$ and $\tilde{H}_2$: nearest-neighbor correlators $C^b=s_i s_j$ and triplet correlators $C^t=s_i s_j s_k$. Here, the index $b$ denotes the bond between site $i$ and $j$ and $t$ marks the triplet defined by $i,j,k$. The Huse-Elser wavefunction can be recovered by choosing the biases on $C^b$ and $C_t$ appropriately. In particular, the optimal parameters found in \cite{huse1988simple} correspond approximately to the biases
\begin{align*}
a^b_{i,j}=-\frac{1.45}{8}, \ \ 
a^t_{i,j,k}=\frac{0.13}{8}\gamma_{ijk} \bf{i},
\end{align*}
where $a^b_{i,j}$ and $a^t_{i,j,k}$ correspond to the biases related to the correlators $C^b$ and $C^t$. We found the influence of the triplet correlators to occur mainly in the biases and therefore set all weights connecting triplet correlators to hidden neurons equal to zero in order to reduce parameters.
The relative error of the variational energies obtained with such a cRBM are compared with the energies of an RBM ansatz in Fig.~\ref{fig:L5}.  For the same amount of parameters, the cRBM ansatz provides significantly lower energies. In addition, the slope of the relative error curves with increasing number of parameters is steeper for the cRBM, indicating that the extension of the spanned Hilbert sub-space is more efficient when introducing correlators.





\bibliographystyle{unsrt}
\bibliography{papers.bib}

\end{document}